%% file: metals.tex
\documentclass[numberedappendix, iop, apj]{emulateapj}
\usepackage{apjfonts}
\usepackage{natbib,natbibspacing}
\usepackage{subfigure}
\usepackage{appendix}

\usepackage{url}

\usepackage{etex}
\reserveinserts{18}
\usepackage{morefloats}

\def\lya{Ly-$\alpha$}

\def\nh{$N_{\rm H I}$}
\def\Z{$\langle {\rm Z}\rangle$}
\def\mh{[M/H]}
\newcommand{\kms}{km~s$^{-1}$ }

\newcommand{\noprint}[1]{}
\newcommand{\figsetstart}{{\bf Fig. Set} }

\newcommand{\figsetnum}[1]{{\bf #1.}}
\newcommand{\figsettitle}[1]{ {\bf #1} }

 \slugcomment{Accepted by ApJ May 21, 2012}

\shorttitle{Metallicity Evolution of DLAs out to $z\sim5$}
\shortauthors{Rafelski, Wolfe, Prochaska, Neeleman, Mendez}

\begin{document}

\title{Metallicity Evolution of Damped {\lya} Systems out to {$\lowercase{z}\sim5$}}

\author{Marc Rafelski\altaffilmark{1}, Arthur M. Wolfe\altaffilmark{2}, J. Xavier Prochaska\altaffilmark{3}, 
Marcel Neeleman\altaffilmark{2}, Alexander J. Mendez\altaffilmark{2}}

\altaffiltext{1}{Infrared Processing and Analysis Center, Caltech, Pasadena, CA 91125, USA}
\altaffiltext{2}{Department of Physics and Center for Astrophysics and Space Sciences, UCSD, La Jolla, CA 92093, USA}
\altaffiltext{3}{Department of Astronomy \& Astrophysics,  UCO/Lick Observatory, 1156 High Street, University of California, Santa Cruz, CA 95064}

\begin{abstract} 
We present chemical abundance measurements for 47 damped {\lya} systems (DLAs), 30 at $z>4$,
observed with the Echellette Spectrograph and Imager and the High Resolution Echelle Spectrometer on the Keck telescopes. 
\ion{H}{1} column densities of the DLAs are measured with Voigt profile fits to the {\lya} profiles, 
and we find an increased number of false DLA identifications with SDSS at $z>4$ 
due to the increased density of the Lyman-$\alpha$ forest.
Ionic column densities are determined using the apparent optical depth method,
and we combine our new metallicity measurements with 195 from previous surveys to determine the evolution of the 
cosmic metallicity of  neutral gas. 
We find the metallicity of DLAs decreases with increasing redshift, 
improving the significance of the trend and extending it to higher redshifts, with a 
linear fit of $-0.22\pm0.03$ dex per unit redshift from $z=0.09-5.06$.
The metallicity ``floor'' of $\approx 1/600$ solar continues out to $z\sim5$, despite our sensitivity for finding DLAs with
much lower metallicities. However, this floor is not statistically different from a steep tail to the distribution.
We also find that the intrinsic scatter of metallicity among DLAs of $\sim0.5$ dex continues out to $z\sim5$.
In addition, the metallicity distribution and the $\alpha$/Fe ratios of $z>2$ DLAs
are consistent with being drawn from the same parent population with those of halo stars. 
It is therefore possible that the halo stars in the Milky Way formed out of gas that commonly exhibits DLA absorption at $z>2$. 

\end{abstract}

\keywords{galaxies: abundances --- galaxies: ISM --- galaxies: evolution --- galaxies: general --- Galaxy: halo --- quasars: absorption lines} 

\clearpage

\section{Introduction}

The recent detection of star-forming galaxies out to $z\sim8$ \citep{Bouwens:2010, Bunker:2010}
provides new information about the crucial  formative stages of galaxy evolution. 
One of the important goals of these studies is to identify the high-redshift progenitors
of current galaxies and their stellar populations. 
A valuable diagnostic tool that links the properties of high-redshift galaxies to their descendants is
their chemical properties \citep{Pettini:2006}. While chemical properties are
exceedingly difficult to extract from the faint starlight emitted by star-forming galaxies
at high redshift, they are more easily determined from protogalactic gas detected in absorption against
bright background quasars \citep[e.g.][]{Pettini:2006, Prochaska:2003a}.
If the metallicity of such  gas were to resemble the metallicity of, say, globular clusters,
then a reasonable first guess is that stars in globular clusters form out of this gas. 
The evolution of the metallicity and other properties of this gas could then provide information
about the sequence of events leading to the formation of current galaxies.

The high-redshift Universe is occupied by populations of objects spanning a wide range of metallicities. While the 
mean metal abundances of dense, compact objects such as quasars is given by [M/H]\footnote{We adopt 
the standard notation in which  the metal abundance [M/H] signifies the logarithmic abundance of 
element M relative to solar; i.e.,  [M/H]$\equiv$~log$_{10}$(M/H)$-$log$_{10}$(M/H)$_{\odot}$. }
$\approx$ 0 \citep{Hamann:1999}, diffuse large-scale configurations such as the {\lya} forest clouds
have [M/H]=$-$3 \citep{Aguirre:2004}. 
In fact, observers have now even identified gas without any trace of heavy elements \citep{Fumagalli:2011hx}. 
With a mean metal abundance of [M/H]=$-1.4$ \citep{Pettini:2006},
the damped {\lya} absorption systems \citep[DLAs; see ][for a review]{Wolfe:2005} are galactic-scale
objects that lie between these two extremes. The purpose of this paper is to present the results of
a new survey for the metal abundances of DLAs out to $z\approx5$. We focus on DLAs for several reasons.
First, since they are detected in absorption, DLAs are unbiassed with
respect to luminosity and presumably mass.
Second, in contrast to all other classes of quasar absorption systems, the gas in DLAs is 
mainly neutral. In fact, DLAs dominate the neutral-gas content of the Universe out to $z\approx5$, 
\citep{Prochaska:2005, Omeara:2007}, and exhibit properties indicating that they are neutral-gas reservoirs for star formation at high
redshift \citep{Nagamine:2004a, Nagamine:2004b, Wolfe:2006}. Moreover, the large optical depth at the Lyman limit of the neutral gas in DLAs 
(typically $\tau_{\rm LL}$ $\approx$ 10$^{4}$) eliminates the need for uncertain ionization corrections  to deduce the metal 
abundances. Accordingly, the most accurately determined  abundances at high redshift are for DLAs \citep{Wolfe:2005}.
Third, because DLA gas likely originates in the outer regions of high-$z$ galaxies \citep[e.g.][]{Rafelski:2011},  
chemical and dynamical information would be preserved due to the long time-scales for dissipation and dynamical
mixing \citep{Freeman:2002}.  

Our primary goal is to measure the cosmic metallicity,
 $\langle {\rm Z}\rangle$~$\equiv$~log$_{10}$(${\Omega_{\rm metals}}/{\Omega_{\rm gas}}$) 
 $-$ ${\rm log}_{10}{\rm (M/H)}_{\odot}$ \citep{Lanzetta:1995},
 at redshifts $z>4$. Previously
we determined {\Z} by measuring the column-density
weighted mean metal abundances \citep[see][]{Prochaska:2003b} 
in redshift bins containing equal numbers of DLAs.
The data revealed a statistically significant  decline in {\Z}
with increasing redshift, given by $\langle {\rm Z}\rangle=(-0.26 {\pm} 0.07)z-(0.59 {\pm}0.18)$, 
which was confirmed at $z$ $<$ 4 by independent measurements \citep{Kulkarni:2005, Kulkarni:2007,Kulkarni:2010}.
While the evolution of {\Z} appears linear in redshift, it is non-linear
in time, with the slope, d{\Z}/dt, becoming significantly steeper at earlier cosmic times.
Since the \citet{Prochaska:2003b} sample contained only 10 DLAs with $z$ $>$ 4, {\Z}, and its derivatives
were poorly determined at such high redshifts. Therefore, we aim to determine whether this
trend continues back to $z$ $\approx$ 5. \citet{Prochaska:2003b} also reported (1) a ``metallicity
floor'' at [M/H]=$-$3.0, because no metal abundances were found below this value despite their sensitivity to [M/H] values
well below $-$4.0,  and (2) a large dispersion of [M/H] at all redshifts.
We wish to investigate whether these phenomena persist at higher redshifts,
as at some point in time the gas should be pristine \citep{Fumagalli:2011hx}.
Although increasing the redshift range of our survey to $z$ $\approx$ 5 corresponds to an increase in survey
time interval of only $\approx$ 0.5 Gyr, we shall show that this is a critical time interval in which, as stated above, 
d{\Z}/dt becomes increasingly steep with decreasing t. 

In addition, we wish to assess if there is a systematic bias in the Sloan Digital Sky Survey (SDSS) {\nh} statistics at $z>4$ \citep{Prochaska:2009, Noterdaeme:2009}, 
such as the column density distribution function, total covering fraction, and cosmological {\nh} mass density. 
These statistics are determined solely from the discovery spectra of the SDSS database, with
low resolution of FWHM$\sim2$\AA ~and sometimes at low signal-to-noise (S/N) (with a cut at S/N$>4$). The analysis of the damped Ly-$\alpha$ profile is significantly affected
by line-blending with the Ly-$\alpha$ forest `clouds',  and the increase of the density of the Ly-$\alpha$ forest at high redshift may lead to an overestimate of 
the {\nh} measurements and an increased false positive identification rate for DLAs. While such biases are likely to be small at $2\lesssim z \lesssim 4$
\citep{Prochaska:2005}, our new measurements resolve out the Ly-$\alpha$ forest and enable the constraint of such a bias at $z>4$. 

Another goal is to compare the metallicity distribution of DLAs with those of known stellar populations. 
Previous studies \citep{Pettini:1997, Pettini:2004, Pettini:2006} have shown that the distribution of metallicities at $z>2$
disagree with known stellar populations in the Galaxy. 
However, if high redshift galaxies are surrounded by a thick disk of neutral gas 
\citep{Prochaska:1997b, Wolfe:2003b, Wolfe:2008, Rafelski:2009, Rafelski:2011, Fumagalli:2011},
then stars that formed {\it in situ} in such gas at high redshift should be present in $z=0$ galaxies, 
such as the Milk Way. These stars would presumably have a similar metallicity distribution as the DLA gas.
We wish to investigate whether or not this disagreement persists in our sample that is larger in size and
extends to higher redshifts.

The outline of this paper is as follows: The observations and data reduction are presented in \S\ref{obsandred}
and the analysis of the resultant spectra is detailed in \S\ref{analysis}. 
In \S\ref{voigt}, Voigt profiles are fitted to the damped {\lya} profiles to determine the H~I
column densities, {\nh}. We discuss the possible elemental abundances that can be used to 
determine the metallicity of DLAs in \S\ref{order}. 
Column densities of the metal lines are determined from the apparent optical
depth method \citep{Savage:1991, Prochaska:2001b}, and combined with the {\nh} values 
to determine element abundances for the DLAs in \S\ref{abund}. This sample is augmented by unbiased
abundances obtained from the literature as described in \S\ref{lit}. We justify the use of various elements for determining
the metallicities of DLAs in \S\ref{comp}.
The determination of {\Z} as function of redshift and time is presented in \S\ref{cosmic},
and a comparison of the DLA metallicity distribution with that of halo stars is shown in \S\ref{halo}. 
In \S\ref{discus} we discuss the metallicity distribution and evolution, and its implications for halo star formation.
Specifically, we compare recent simulations to our data and discuss the metallicity evolution in \S \ref{discuss_metalevol}, 
and consider the implications of the metallicities and $\alpha$-enhancement of DLAs for halo star formation in 
\S \ref{halocomp}. In \S\ref{sum} we summarize the results.

Throughout this paper we adopt a  cosmology with
(${\Omega_{\rm M}},{\Omega_{\Lambda}},h$)=(0.3,0.7.0.7).

\section{Observations and Data Reduction}
\label{obsandred}

The primary scientific goals of our survey are to: 
(i) perform an
unbiased evaluation of the metallicity in neutral atomic-dominated hydrogen gas at $z>4$; 
(ii) confirm and refine previous estimates on the incidence and
\ion{H}{1} mass density in $z>4$ DLAs; and
(iii) identify new targets for follow-up ALMA observations in the
\ion{C}{2}* 158\,micron transition.
These goals were well suited to the spectroscopic characteristics of
the Echellette Spectrograph and Imager \citep[ESI;][]{Sheinis:2002}, with the possibility of
followup observations with the High Resolution Echelle 
Spectrometer \citep[HIRES;][]{Vogt:1994} on the Keck I and II 10m telescopes.
We therefore required a set of $z>4$ DLAs that could be observed from Keck observatory.    

The first sample of targets were drawn from the SDSS-DR3 and SDSS-DR5 surveys for damped
\lya\ systems performed by \cite{Prochaska:2005} and \cite{Prochaska:2009} respectively.  
We restricted the list to sources that were sufficiently bright such that we could obtain
S/N~$>15$ spectra in a non-prohibitive observing time.  
After the first few nights, we needed to supplement the sample with
additional $z>4$ DLAs, and we searched for new $z>4$ quasars from
SDSS-DR7. 
These additional targets will not be included in the \ion{H}{1} analysis as
they were selected in a more heterogeneous manner than the original
set.  They are unbiased, however, with respect to the gas metallicity because they were selected solely on the presence of strong \ion{H}{1} Ly$-\alpha$ absorption.
In addition to this primary high redshift quasar sample obtained in 2009-2011, we also include new observations of lower 
redshift quasars ($2.5 \lesssim z \lesssim 3.5$) obtained in 2008 to augment the number of unbiased metallicity measurements (see below). 
We note that the effective upper limit to the redshifts of our DLA sample is about 5.3; DLAs with higher redshifts are essentially undetectable
because line blanketing by the {\lya} forest severely attenuates the quasar continuum at wavelengths shortward of {\lya} emission;
i.e. at wavelengths where such DLAs would be located \citep{Fan:2006, Songaila:2010}

The new observations comprise a total of 6 nights (effectively 5 nights when considering the losses to poor weather) on ESI and 7 nights on HIRES. 
In Table \ref{metal:tab1} we present a journal of observations describing the
50 quasars in our sample. Column (1) gives the quasar coordinate name obtained from the SDSS
survey. Columns (2) and (3) give the RA and Dec. of the quasar in J2000. 
Columns (4) and (5) give the AB $r$ and $i$ magnitudes
of the quasar: in most cases the $i$ magnitude gives
a realistic measure of the unattenuated quasar brightness since at 
the redshifts of most of the DLAs, 
the $i$ filter is placed redward of {\lya} emission.
Column (6) gives the emission redshift of the quasars,
and column (7) lists candidate DLA absorption redshifts obtained from SDSS.  
Column (8) provides the UT date of the observations, 
column (9) states the instrument used, and column (10) lists the slit size. 
Column (11) gives the exposure time, and column (12) the S/N pixel$^{-1}$ of the data. 

\input{tab1.tex}

In general, we obtained 2-3 exposures per quasar with ESI, with sufficient exposure time to reach S/N values 
larger than 15 pixel$^{-1}$ (11 \kms). Observations were reduced on the fly to 
determine if the DLA candidate did not meet our threshold criteria for 
being a DLA, i.e. N(HI)$\geq 2\times10^{20}$cm$^{-2}$. If it did not,
 then we abandoned obtaining more exposures, resulting in reduced S/N values for those targets. 
These quasars are nonetheless listed in Table \ref{metal:tab1}, and the corresponding DLAs are
marked with the footnote {\it b}. Table \ref{metal:tab1} also lists the median S/N per pixel at 
$\lambda\approx 8300$\AA\footnote{The HIRES observations of J1353+5328 and J1541+3153 have their S/N determined at $\lambda\approx 7300$\AA ~due to their lower redshifts and corresponding HIRES coverage.},
corresponding to the typical wavelength of the measured metal absorption lines. The average S/N of the ESI 
spectra with confirmed DLAs is 25, with a minimum S/N of 14. The mean S/N of the spectra with DLA candidates not meeting our criteria is 15, with a minimum S/N of 9. 

All ESI observations prior to 2010, and all HIRES observations 
were observed at the same slit position to facilitate the removal of cosmic rays between exposures.
In order to decrease the effects of fringing in ESI observations \citep[$\sim15\%$ at 9000\AA;][]{Sheinis:2002}
at the longest wavelengths ($\lambda > 8,000$\AA), we used a three point dither pattern along the slit
for all ESI observations starting in 2010.
Throughout the ESI observations, we primarily used the 0.$\tt''$75
slit, except in two cases when we used the 0.$\tt''$5 slit, which correspond to 
a resolution of FWHM of $\sim44$ and $\sim33$ km s$^{-1}$  
respectively ($R\approx 7,000-10,000$).
For the HIRES observations, we used the C1 decker, with a 0.$\tt''$86 slit corresponding 
to a resolution of  FWHM of 6.2 km s$^{-1}$ or 1.5 km s$^{-1}$ pix$^{-1}$  ($R\approx50,000$).
ESI has a fixed format echellette that covers the spectral region $\lambda=$ 4,000 to 11,000\AA,
while the HIRES echellette coverage varies with each setup. The HIRES observations are all
followup observations of the ESI observations in order to deblend lines, check for saturation,
and measure weak lines, and we therefore targeted the appropriate
metal lines for metallicity determination with the HIRES setups. 

\begin{figure*}[h!]
\center{
\includegraphics[scale=0.9, viewport=0 5 550 700,clip]{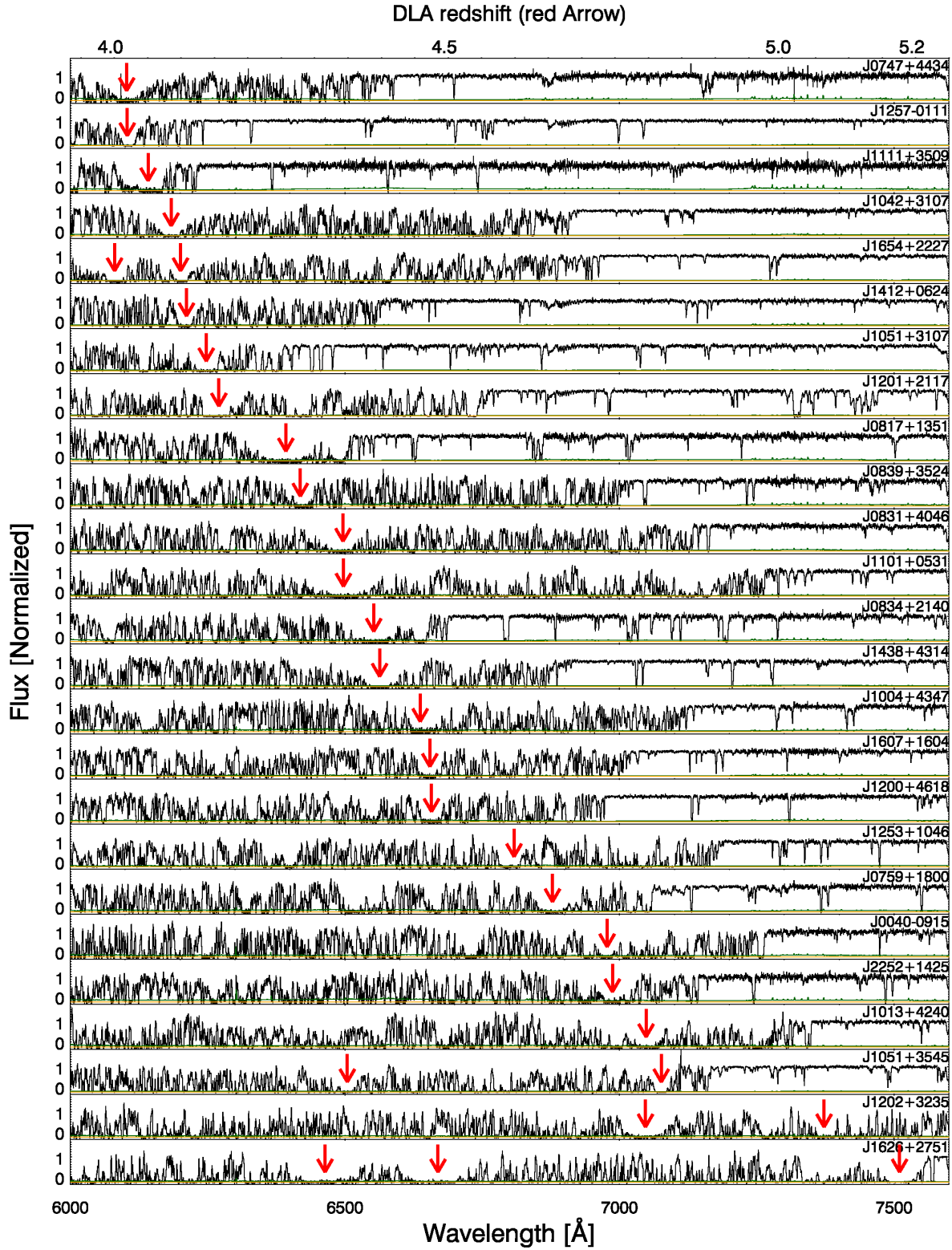}
}
\caption{
ESI spectra of 25 quasars with 30 DLAs at $z>4$. 
Fluxes (black curves) and 1-$\sigma$ error arrays (green curves) are
normalized to unit continuum. The orange lines are zero flux.  
$z>4$ Damped Ly$\alpha$ lines, denoted
by red arrows, are listed in order of increasing 
value of the largest redshift DLA for each spectrum
}
\label{fig:multi}
\end{figure*}

Figure \ref{fig:multi} shows the 25
quasar spectra containing 32 confirmed DLAs with $z > 4$ (see \S\ref{voigt} regarding confirmation of DLAs), 
in order of increasing value of the largest redshift DLA
for each spectrum. We note that while these spectra also contain other lower redshift 
DLAs, we only show those with $z > 4$. 
In addition to the 25 quasars containing $z > 4$ DLAs, we also analyze ESI observations and HIRES followup from 2008
which were obtained in a similar unbiased fashion.

All the data were reduced with the $\tt XIDL$\footnote{See \url{http://www.ucolick.org/~xavier/IDL/}}
package developed in IDL by J. X. Prochaska. 
Specifically, the ESI observations were reduced with $\tt ESIRedux$\footnote{See
\url{http://www2.keck.hawaii.edu/inst/esi/ESIRedux/}} \citep{Prochaska:2003a} 
and the HIRES observations were reduced with $\tt HIRedux$\footnote{See 
\url{http://www.ucolick.org/~xavier/HIRedux/}}, both of which are publicly available. 
All the data were extracted with optimal extraction, and the data were continuum fit
using the $\tt x\_continuum$ routine within $\tt XIDL$.

\section{Analysis and Results}
\label{analysis}
In this section we present the analysis of the quasar spectra obtained  in this survey. 
We first confirm the DLA candidates using Voigt profile fitting,  determine their {\nh} values, and 
investigate the false positive rate of SDSS-selected DLAs at high redshift. 
We discuss the possible elemental abundances that can be used to 
determine the metallicity of DLAs, and then describe the derivation of elemental abundances.
We then compile the latest literature metallicities, 
justify the use of multiple elements for metallicity determination, and determine {\Z} as function of redshift. Finally, we compare the metallicity 
distribution of DLAs to that of known stellar populations in the Milk Way, focusing on the halo stars.

\subsection{\ion{H}{1} Column Densities}
\label{voigt}
We determine the {\nh} values of the damped {\lya} systems in our sample using the 
same methodology as \citet{Prochaska:2003a}. Specifically, we fit the {\lya} line
with a Voigt profile in the fluxed, but not continuum fit, spectra using the routine
$\tt x\_fitdla$, part of the $\tt XIDL$ package. We tie the centroid of the 
Voigt profile to the redshift of the low-ion metal-line
transitions\footnote{All of the bona-fide DLAs exhibit at least one positive detection.}, which yields a velocity centroid 
to within $\sim30$ km s$^{-1}$ of the true line center \citep{Prochaska:2003b}. We simultaneously fit the Voigt profile 
and the continuum to obtain a fit that matches both the core and wings of the profile. The uncertainties associated with 
the redshift errors are significantly less than the uncertainty in the continuum fit, and  therefore are not  
separately included in the fitting procedure. 
While this fitting procedure is not completely quantitative, it is the standard methodology 
and is justified by the fact that the uncertainties are dominated by systematic errors: continuum fitting and line-blending. 
In our experience, a standard $\chi^2$ analysis yields unrealistically small error 
estimates \citep{Prochaska:2003a}. 
We place very conservative error estimates on the {\nh} values by manually selecting values that ensure 
that the error region encompasses all possible solutions for {\nh}. These values systematically increase with increasing redshift 
due to the increased {\lya} forest blending, and the reader should consider them to be roughly at the 95\% confidence level.

The sample of candidate DLAs all have {\nh}$\geq 2\times10^{20}$cm$^{-2}$ based on the SDSS
spectra with spectral resolution FWHM $\sim2$\AA. At low redshift, this is sufficiently high resolution to obtain relatively 
accurate {\nh}, and therefore yields only a small number of false candidates.  However, as we move to higher redshift, the 
{\lya} forest becomes significantly more dense, and the damped {\lya} profiles in the SDSS 
spectra are significantly affected by line-blending with random {\lya} forest clouds. 
Since the ESI spectra resolve out much of the {\lya} forest, we can therefore obtain more robust values of N$_{\rm H I}$. 
In order to confirm a candidate DLA from SDSS to be a true DLA, we require {\nh}$\geq 2\times10^{20}$cm$^{-2}$ 
in the ESI spectrum. 

We fitted {\nh} Voigt profiles to 68 candidate DLAs from SDSS, and confirm 51 as bona-fide DLAs. 
Figures \ref{fig:rogue0} and \ref{fig:rogue1} show the Voigt profile fits of 51 newly confirmed DLAs 
ordered by increasing RA. The velocity centroid of the DLA is determined from the low-ion metal transitions, 
and some panels have more than one DLA candidate marked, and in those cases the central DLA at zero velocity 
is the DLA for the information given in that panel. The velocity interval of the individual 
panels varies with column density such that the wings of the damped profile are presented.

The higher density of the {\lya} forest at higher redshifts results in more line blending than at lower redshift, and therefore sometimes 
one or both of the wings are not clearly defined, and in these cases we rely on the core of the line. In a 
few other cases, a strong {\lya} absorber is blended with one side of the DLA's core, and 
the resultant profile is constrained by the core on the other side, and the wings on one or both sides. 
For example, the Voigt profile fit of J1200+4618 at $z_{\rm abs}=4.477$ turns up more sharply than the data on the blue side of line center due to confusion with one or more {\lya} absorbers at $v\sim-900$ km s$^{-1}$. The same effect can be seen in J1201+2117 at $z_{\rm abs}=4.158$, and a few other DLA fits.
These challenges are not a problem, as the redshift is determined from the low-ion metal lines, and we therefore have 
sufficient information for a robust determination of {\nh}.

The 17 false candidates consist of multiple lower column density lines that are blended together in the lower resolution SDSS data. 
Of the 68 candidate DLAs in Table \ref{metal:tab1}, 
46 of them have $z > 4$. Of those 46 candidate $z > 4$ DLAs, 14 are false positives, leaving a total of 32 confirmed
DLAs with $z > 4$. As a result, we find that $\sim$30\% of $z>4$ DLAs are misidentified in SDSS due to line-blending.
In contrast, for our sample of 22 $z<4$ DLAs, we find 3 false candidates, none of which are one of the 
14 $z\sim3$ candidate DLAs obtained in our 2008 run. To investigate this further, we limit ourselves to the more
homogeneous part of the sample that were identified with the automated routine from DR5 SDSS data \citep{Prochaska:2009}, 
excluding the manually selected DR7 SDSS DLAs (see \S \ref{obsandred}).
Also, we exclude 2 DLAs that were lower than {\nh}$\geq
2\times10^{20}$cm$^{-2}$ in 
the SDSS data, but were measured in the ESI spectra
to meet the DLA criterion.
This removes 5 false candidate DLAs (4 of which are at $z>4$) and 9 confirmed DLAs from the sample, leaving a total of 56 candidate DLAs. 
Out of these 38 candidate $z > 4$ DLAs, 10 are false positives, leaving a total of 28 confirmed DLAs with $z > 4$, and a $\sim$26\% misidentification rate.
For the 18  $z<4$ DLAs, we find 2 false candidates, and a $\sim$11\%
misidentification rate. 
We note that one of these two DLAs is found in a quasar with  $z_{em}=4.9$,
 and therefore is severely affected by the $z>4$ Ly$-\beta$ forest, which is uncommon, and therefore the misidentification rate at $z<4$ is likely closer to $\sim$6\%. 
Despite the small number statistics, there clearly is a larger percentage of false positives at $z>4$ than at $z<4$.

We compare the {\nh} values determined from the DR5 SDSS data and 
from our ESI data in Figure \ref{fig:nhi}. The top panel is the ESI {\nh} versus the SDSS {\nh}, while the bottom
panel are number counts of misidentified DLAs.
The dotted line marks the minimum criteria to be a DLA,  {\nh}$\geq 2\times10^{20}$cm$^{-2}$. There is general good agreement
between the ESI  {\nh} values and the SDSS  {\nh} values within their uncertainties. In addition, the histogram shows that
the primary misidentifications occur for DLAs near the {\nh}$=2\times10^{20}$cm$^{-2}$ limit. Some of these misidentified systems 
are super Lyman limit systems, although the majority are due to a combination of lines in the  Lyman-$\alpha$ forest. 
Studies using DLA identifications from SDSS at $z>4$ \citep{Prochaska:2009} therefore need to 
apply a correction for the increased false positive rate before determining the column density distribution function, the total 
covering fraction, and the integrated mass density of \ion{H}{1} gas. The histogram in Figure \ref{fig:nhi} suggests that this correction 
mainly needs to be applied for the lowest {\nh} DLAs.

\begin{figure*}
\center{
\includegraphics[scale=0.9, viewport=30 15 520 620,clip]{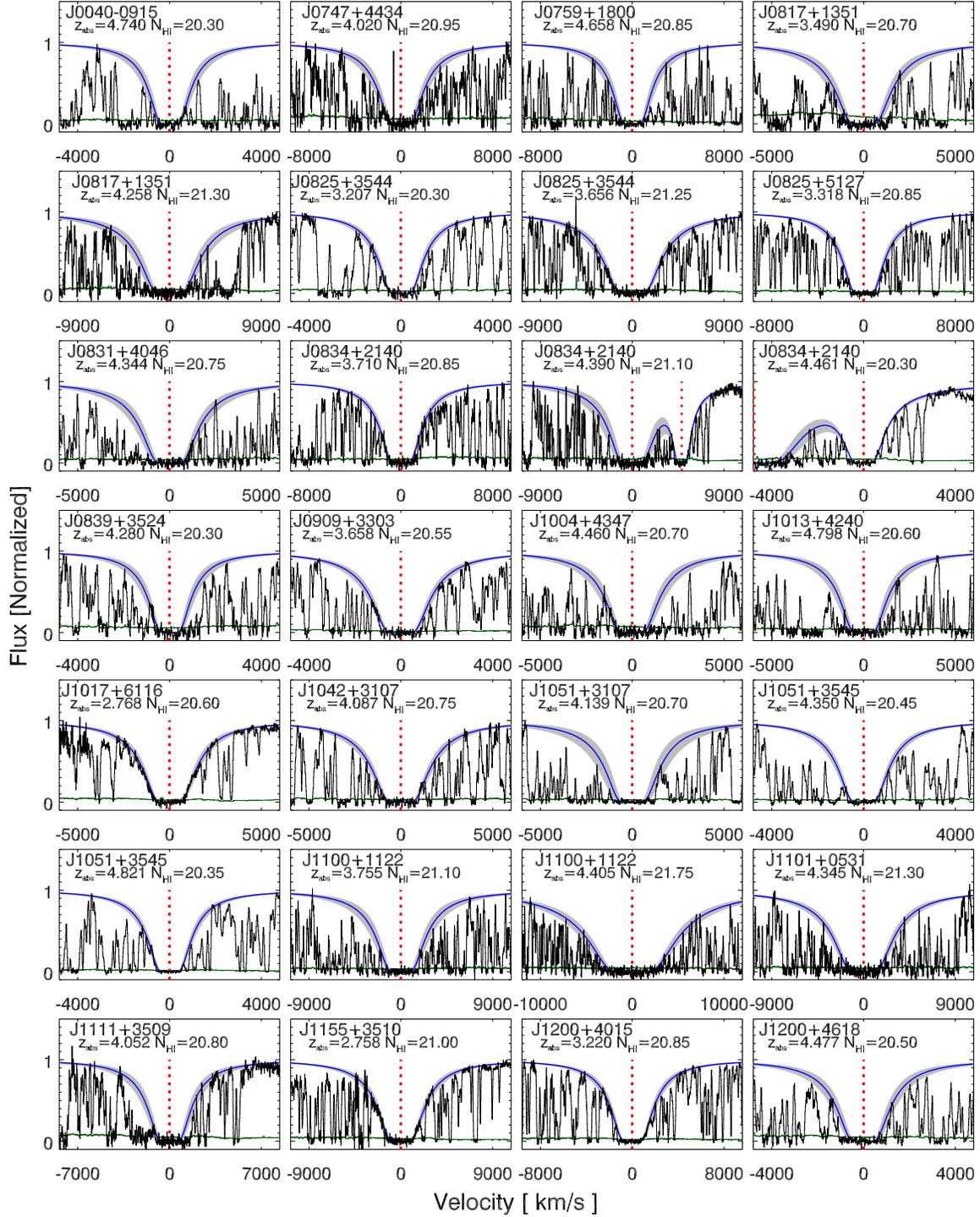}
}
\caption{Voigt profile fits to 51 confirmed DLAs ordered by RA, with the x-range based on the {\nh}. Blue curves are best-fit 
profiles and gray shade includes 95\% confidence limits surrounding best fits.
The red dotted line marks the velocity centroid of the DLA determined from the low-ion metal transitions, with possible multiple components 
as determined by the metal lines. 
Individual panels include quasar name, DLA redshift and H I column density.
The green line represents the uncertainty.
}
\label{fig:rogue0}
\end{figure*}
\begin{figure*}
\center{
\includegraphics[scale=0.9, viewport=30 100 520 620,clip]{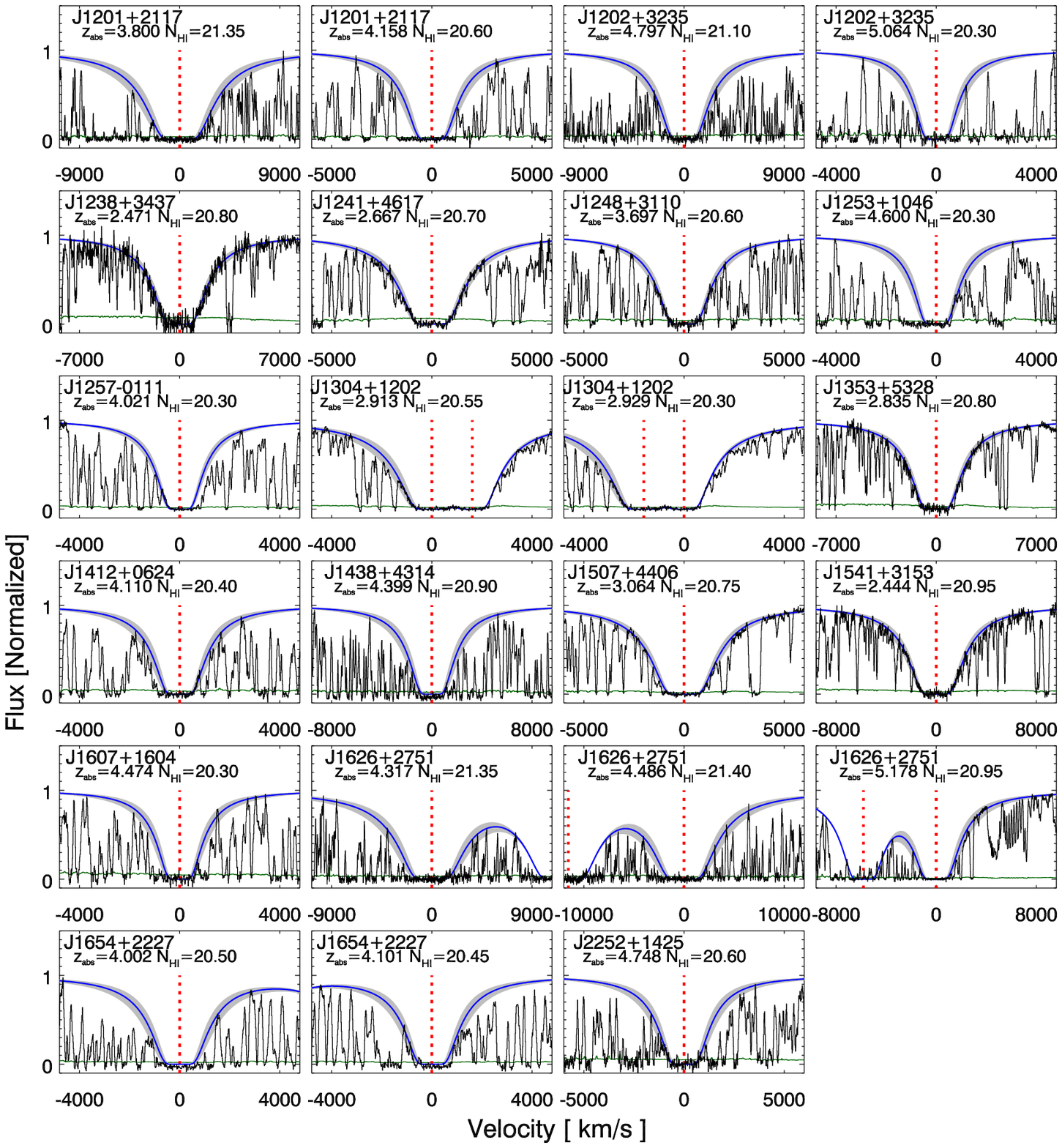}
}
\caption{Continuation of Figure \ref{fig:rogue0}.
}
\label{fig:rogue1}
\end{figure*}

For such future corrections, we break down the $\sim$26\% misidentification rate into random and systematic misidentifications. Of the 10 false positives from DR5,
3 are random misidentifications with high {\nh} absorbers at the correct redshift that do not quite meet the criteria of {\nh}$\geq 2\times10^{20}$cm$^{-2}$, 
but are within $1-2\sigma$ of their previous measurements. The other 7 false positives are systematic misidentifications due to a combination of many lower {\nh} absorbers 
masquerading as a single higher {\nh} absorber. Figure \ref{fig:roguebad} show Voigt profile fits to both the original SDSS data and the higher resolution ESI data for 
the 7 systematic misidentifications from DR5.

We expect that a number of absorbers with {\nh}$<2\times10^{20}$cm$^{-2}$ will randomly be measured above this value,
and that a number of absorbers with {\nh}$>2\times10^{20}$cm$^{-2}$ to be measured below this value. We check if the number of random misidentifications measured
is as expected from the uncertainties and the column density distribution function using the Eddington bias formalism \citep{Eddington:1913, Teerikorpi:2004}.  
We limit ourselves to DLAs with {\nh}$\lesssim20.5$,  as above this {\nh} the problem is small (see Figure \ref{fig:nhi}) and a random error is unlikely to move an absorber with
{\nh}$<2\times10^{20}$cm$^{-2}$ to a higher {\nh}. The observed number of DR5 DLAs with $z>4$ and {\nh}$<20.5$ is 9, to which we add the 3 random misidentifications for a total of 12 absorbers.
We use an uncertainty of 0.2 dex, and model the column density distribution function. The expected number of DLAs from the Eddington bias formalism is 8.4, resulting in an observed difference of 3.6 DLAs. This is consistent with the 3 random misidentifications measured, especially given the 
small number statistics considered here. 

We note that in our survey for DLAs, we reject proximate DLA systems 
that are found to be within 3000 km s$^{-1}$ of the
quasar emission redshift, as these DLAs may be physically associated or otherwise 
affected by the quasar \citep[see][]{Prochaska:2008, Ellison:2010,
  Ellison:2011}. This removes our highest redshift confirmed DLA ($z=5.18$, J1626+275), and it is not included in the analysis of the metallicity evolution, although we include it for completeness in our tables.

\begin{figure}
\center{
\includegraphics[scale=0.45, viewport=0 5 495 360,clip]{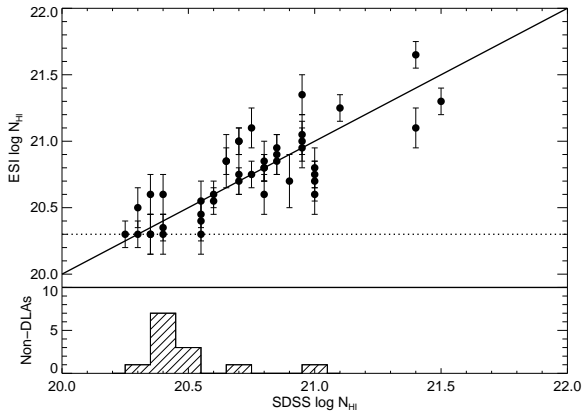}
}
\caption{Comparison of the {\nh} values determined from the DR5 SDSS data and 
from our ESI data in Figure \ref{fig:rogue0}. The top panel is the ESI {\nh} vs. the SDSS {\nh}, while the bottom
panel are number counts of misidentified DLAs, based on the higher resolution and higher S/N ESI data. 
The dotted line marks the minimum criterion to be a DLA,  {\nh}$\geq 2\times10^{20}$cm$^{-2}$. 
}
\label{fig:nhi}
\end{figure}

\begin{figure*}
\center{
\includegraphics[scale=0.7, viewport=0 120 550 700,clip]{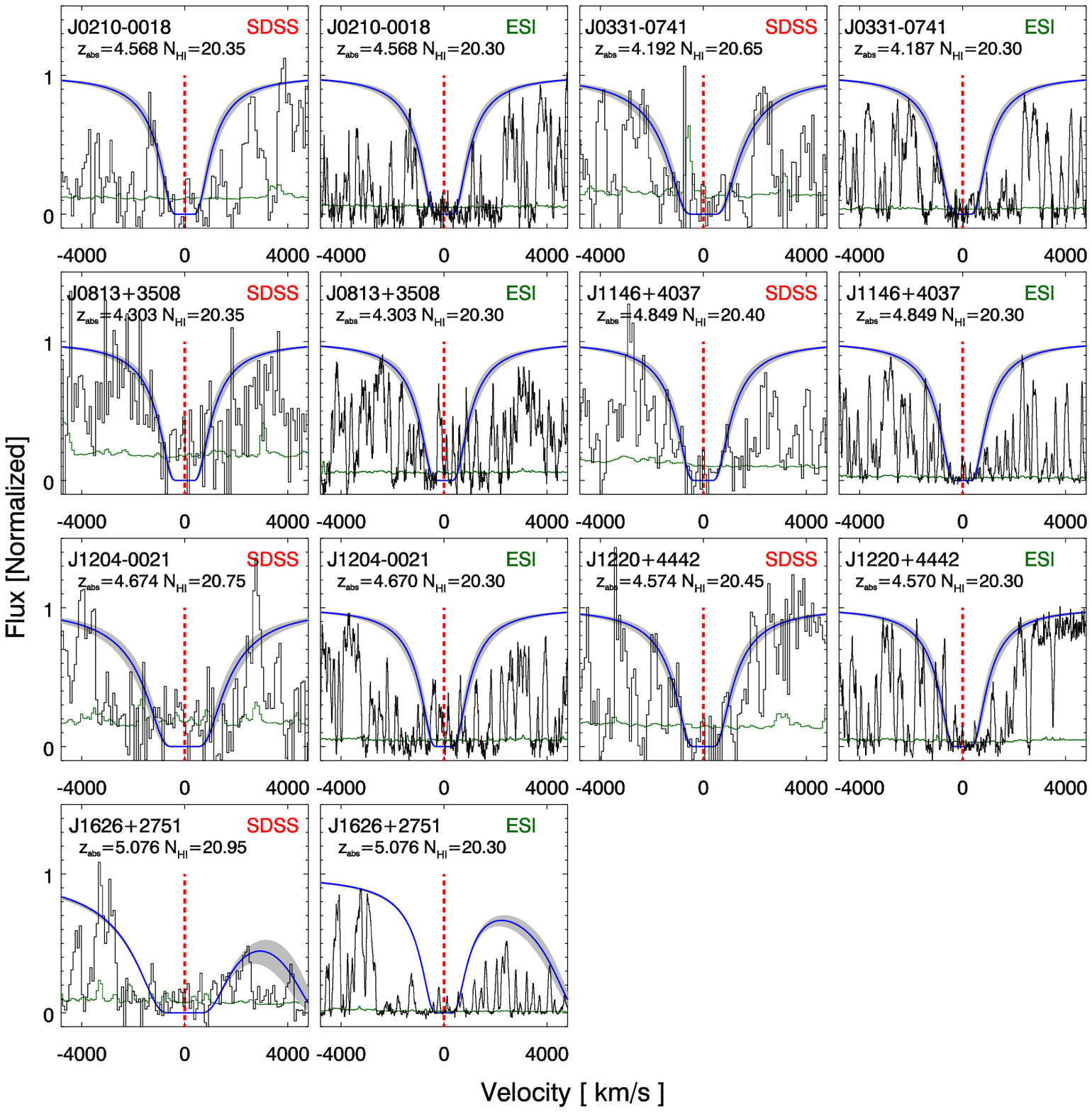}
}
\caption{Voigt profile fits to the 7 systematic misidentifications of DLAs from SDSS. We alternate panels showing the SDSS data and the ESI data,
and label each panel accordingly. 
The blue curves are best-fit profiles to the SDSS data from \cite{Prochaska:2005} and \cite{Prochaska:2009}, or the ESI data from this work.
Individual panels include quasar name, DLA redshift and H I column density.
The gray shade includes 95\% confidence limits surrounding best fits, the red dotted line marks the velocity 
centroid of the DLAs as determined from the low-ion metal transitions, and the green line represents the uncertainty.
}
\label{fig:roguebad}
\end{figure*}

\subsection{Determining DLA Metallicities}
\label{order}

A principal goal of this paper is to trace the build-up of heavy
elements in neutral gas across cosmic time.  In the Sun, oxygen (O)
dominates by number and contributes $\approx 50\%$ of the heavy element
mass budget \citep{Lodders:2010}.  This fraction is likely even higher in
the early universe when Type~Ia supernovae have had less time to 
enrich the iron (Fe) peak elements (see \S\ref{halo}). 
Indeed, O is commonly observed to be enhanced in metal-poor
environments such as our Galactic halo \citep[e.g.][]{Bessell:1991}.  
Ideally, one would assess the heavy element mass density through
measurements of O.  Unfortunately in DLA analysis, the
\ion{O}{1} transitions are generally either too weak or too strong to
permit a precise estimate of its abundance relative to hydrogen.  
After oxygen, carbon (C) and neon (Ne) contribute $\approx 30\%$ of the heavy
element mass in the Sun.  The same challenges as for oxygen apply to
C, and there are no Ne transitions available for DLAs.
It is evident that we are forced to use a proxy for the heavy element
abundance.

The elements nitrogen (N), magnesium (Mg), silicon (Si), sulfur (S), 
and Fe each contribute $\approx 5\%$ of the
heavy element mass density in the Sun and each of these can, in
principle, be measured in DLAs.  To best trace the dominant elements
(O, C, Ne), one would prefer an element that is produced and expelled in
tandem, i.e. predominantly in the supernovae explosions of massive
stars.  Both Si and S are good candidates since they lie along the $\alpha$-element sequence of
nucleosynthesis, are produced in massive stars, and are easily measurable 
in high redshift DLAs. S is a volatile element with negligible depletion and with 
multiple lines of varying oscillator strengths. Alas, these lines are often in the {\lya} forest, making 
them measurable in only a fraction of DLAs. 
Si has many lines of varying oscillator strengths at multiple wavelengths, which results in at least
one line being unsaturated and outside the forest for the majority of DLAs. 
While Si has nearly twice the mass density of S, it is a mildly refractory element and its
gas-phase measurement may underestimate the total metal abundance
\citep{Prochaska:2002b, Vladilo:2011}.
Considering all of these aspects together, we conclude that S is the
best option and use its abundance as a proxy for the heavy element
mass density in DLAs wherever possible\footnote{Aside from the few
cases where oxygen is measured precisely.}.
Under the expectation that metal-poor gas is relatively undepleted, we
adopt Si as the second choice, especially in low metallicity systems. 
We further investigate the possibility of depletion of Si in \S\ref{comp}, 
where we determine that it is not significantly depleted in DLAs based on current measurements.

Within the DLA literature, zinc (Zn) has played a special role in estimating the 
metallicity \citep[e.g.][]{Meyer:1989,Pettini:1990,Pettini:1994, Pettini:1997, Akerman:2005}.  
This reflects both observational convenience and the fact that Zn is non-refractory.  
Specifically, Zn has two strong transitions at 2026 and 2062 {\AA},
that almost always lie outside the {\lya} forest, and are
rarely saturated due to their low oscillator strengths and low Zn abundances.
Unfortunately, Zn is only a trace element which contributes $\approx 10^{-4}$ of the mass density for the heavy elements. 
Furthermore, the weakness of the Zn transitions make it impossible to measure its abundance in low metallicity DLAs, 
and the large rest-frame wavelengths of the transitions make them difficult to measure at high redshift.
In addition,  the nature of Zn is not fully understood;  Zn is not an $\alpha$-element, nor
an Fe peak element. While Zn is often observed to track Fe, theoretical work on the nucleosynthesis of Zn
often predicts it may originate in massive stars \citep{Woosley:1995, Hoffman:1996, Umeda:2002, Fenner:2004}.
On the more positive side, Zn tracks the relative abundances of S
and Si in DLAs, and empirically is a fair substitute for these elements (see \S\ref{comp}).
Therefore, when O, S, and Si are either not covered by our spectra or are saturated, and the Zn transitions are available, 
we use the Zn abundance as a proxy for the heavy element mass density. 

In the small number of cases where none of the above elements are available, we are left with using elements 
such as Fe, nickel (Ni), or aluminum (Al). As Fe and Ni are refractory elements, they will be depleted at high metallicities.
At low metallicities, we can measure the intrinsic abundance of Fe and Ni, which tracks the abundances of S and Si well, 
with an offset due to the $\alpha$-enhancement of DLAs. We determine the $\alpha$-enhancement in \S\ref{abund} and \S\ref{comp} from 
the abundance of Fe combined with an $\alpha$-element abundance.  
We apply an  [$\alpha$/Fe] offset of 0.3 dex, which is slightly different
than that used in some previous studies \citep[0.4; e.g.][]{Prochaska:2002b, Prochaska:2003b}, but is more 
representative of [$\alpha$/Fe] of DLAs \citep[][and see below in \S \ref{lit} and \S\ref{comp}]{Wolfe:2005}\footnote{
We note that the y-axis of Figure 8 in \citet{Wolfe:2005} is mislabeled. The tick labeled 0.5 should be labeled 0.7}.
In addition, we also determine [Fe/H] from iron peak elements for all DLAs, as this is useful for studying the chemistry of
DLAs, and for an estimate of the dust-to-gas ratio, as Fe is a depleted element for DLAs with [M/H] $>-1$ 
\citep{Wolfe:2005}. In cases where Fe is not available, we determine [Fe/H] useing Ni with
a 0.1 dex correction, or Al.
We note that whenever we use  an Fe peak element for the metallicity, we add the uncertainty in [$\alpha$/Fe] of 0.16
to the [Fe/H] uncertainty in quadrature.

\vspace{10mm}

\subsection{Measured Element Abundances}
\label{abund}

In this subsection we describe the derivation of ionic column densities for all of the new confirmed DLAs
using the apparent optical depth method \citep[AODM;][]{Savage:1991}. 
This method can uncover hidden saturated lines by comparing the apparent 
column density per unit velocity, $N_a(v)$, for multiple transitions of a single ion. We then calculate
$N_a(v)$ for each pixel from the optical depth equation, 
and sum over the velocity profile of each transition to obtain the total column density, $N$. 
If multiple lines for a single ion are available, then we take the
weighted mean of the column densities from unsaturated lines. 
The AODM technique yields column densities that agree well with line fitting, 
and it is easier to apply to a large data sets such as this one \citep{Prochaska:2001b}.
We use the wavelengths and oscillator strengths shown in 
Table~2 of \citet{Morton:2003} 
and the meteoritic solar abundances shown
in Table 1 of \citet{Asplund:2009}. 

The medium resolution of ESI could result in possible saturation of unresolved weak metal lines if they were particularly narrow \citep{Penprase:2010}, 
and such lines should normally be treated as lower limits. However, in this study we obtained HIRES followup in order to 
1) check for saturation of narrow weak metal lines from the ESI measurements by resolving the lines, and 
2) obtain metallicities that would otherwise not be measurable with ESI due to blending of lines or weak transitions.
By targeting the most likely problematic candidates with HIRES, we reduce the saturation issues. Specifically, 
we followed up DLAs with narrow lines whose metallicity was based on single metal lines, as these are the most likely culprits.
In the cases where a line looked potentially saturated without available HIRES data, we used a different line instead. In a few cases, this resulted in an 
Fe-element based metallicity rather than an $\alpha$-element based metallicity. 
We therefore use the remaining ESI sample without applying saturation corrections. 
In addition, if we did miss a small number of saturated DLAs, they would not
affect the main conclusions, as we have sufficient measurements per redshift bin such that {\Z} is not sensitive to any 
individual measurement (see \S3.4). Lastly, as we probe higher redshifts, the common element transitions used for metallicity determination
in DLAs are less likely to be saturated due to their lower metallicities at higher redshifts \citep[][and this study]{Prochaska:2003b}.

In Table \ref{metal:tab2} we present new metallicity measurements for 47 DLAs, with 30 of them at $z>4$.
Four DLAs do not have measurable metallicities due to a lack of unsaturated relevant lines 
red ward of the Lyman-$\alpha$ forest, and are not in the table.
We do note, however, that each of these exhibits obvious metal absorption and therefore has a metallicity [M/H] > $-2$.
Another  is a proximate DLA, and is included in the table.
Column (1) gives the quasar coordinate name obtained from the SDSS survey, 
column (2) gives the DLA absorption redshifts determined from the metal transition lines, and 
column (3) gives the logarithm of the measured {\nh} values.
Columns (4) and (5) give the alpha element flag and metallicity, 
and column (6) and (7) give the Fe flag and metallicity. 
Column (8) and (9) give the flag and final metallicity used in this paper. 
The flags in columns (4) and (6) state which transitions are used and if they are detections or limits,
as described in the Table notes. We also provide the measured elemental column densities 
of these DLAs in the online material in the appendix. In addition, we show the velocity profiles for those 
metal-line transitions in the online material. 

\input{tab2.tex}

The uncertainty in [$\alpha$/H],  [Fe/H], and [M/H] in Table \ref{metal:tab2}
include the uncertainty of {\nh} added in quadrature, and therefore
are dominated by the uncertainty in {\nh}. 
We ignore the error in the continuum fit in our error
calculations for metal-line transitions, 
although it may be important for very weak transitions. 
We set a minimum error of 0.1 dex for all metallicity measurements, and note that 
our main results are not sensitive to the uncertainties in individual measurements
(see bootstrap method in \S\ref{cosmic}). 

\begin{figure}
\center{
\includegraphics[scale=0.45, viewport=15 5 495 360,clip]{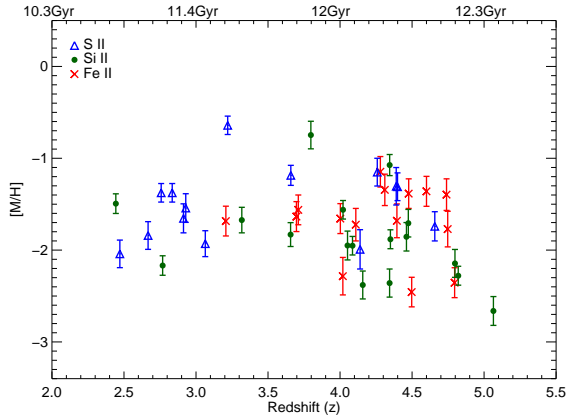}
}
\caption{DLA metal abundance, [M/H], versus redshift for the 46 DLAs in Table \ref{metal:tab2} (not including the proximate DLA). 
The symbols and colors indicate the origin of the [M/H] values: 
\ion{S}{2} measurements (blue triangles)
\ion{Si}{2} measurements (green filled circles), and \ion{Fe}{2} measurements (red crosses). 
We note that the \ion{Fe}{2} [M/H] values are given by [Fe/H]+0.3 dex.
}
\label{fig:metalobs}
\end{figure}

In Figure \ref{fig:metalobs}, we plot the 46 metallicities obtained in this paper as a function of redshift 
(not including the proximate DLA). 
The figure outlines which elements are used to 
obtain [M/H] and the uncertainties of those measurements. 
The symbols and colors indicate the origin of the [M/H] measurements, with blue triangles representing \ion{S}{2}, 
green filled circles representing \ion{Si}{2}, and red plus crosses representing \ion{Fe}{2}.
We note that the metallicities derived from \ion{Fe}{2} include an
$\alpha$-enhancement correction, with [M/H] = [Fe/H]+0.3 dex (see
\S\ref{lit} and \S\ref{comp}).

\subsection{Element Abundances from the Literature}
\label{lit}

In this subsection we describe the compilation of metallicities of DLAs from the literature. 
Over a thousand DLAs have been identified by SDSS \citep[e.g.][]{Prochaska:2005, Prochaska:2009, Noterdaeme:2009},
and many hundreds have been observed with a wide variety of telescopes and instruments with varying resolution, S/N, and
wavelength coverage. Similar to \citet{Prochaska:2003b}, we restrict our analysis to DLAs observed at 
high-resolution ($R>5000$) and high S/N (15 pixel$^{-1}$), and require them to 
meet the DLA criteria of \nh$\geq 2\times10^{20}$cm$^{-2}$. In addition, we avoid samples with metallicity 
biases, such as metal poor samples \citep[e.g.][]{Pettini:2008, Penprase:2010, Cooke:2011} or metal rich samples 
\citep[e.g.][]{HerbertFort:2006, Kaplan:2010}. We make an exception for low redshift DLAs
with $z\lesssim1.5$, as these are generally found based on strong metal lines \citep[e.g.][]{Khare:2004, Kulkarni:2005, Meiring:2006, Peroux:2006}. 
Therefore, other than for DLAs in our lowest redshift bin, the
literature sample is not (explicitly) metallicity biased. 

\begin{figure}
\center{
\includegraphics[scale=0.45, viewport=15 5 495 360,clip]{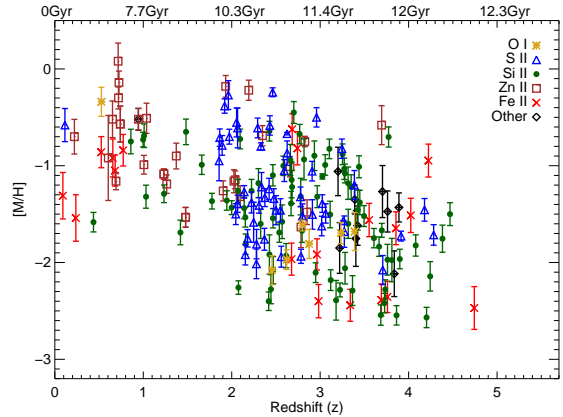}
}
\caption{DLA metal abundance, [M/H],  versus
redshift for 195 DLAs from the literature. The symbols and colors indicate the origin of the [M/H] values: \ion{Si}{2} measurements (green filled circles),
\ion{S}{2} measurements (blue triangles), \ion{Zn}{2} measurements (brown boxes), 
\ion{Fe}{2} measurements (red crosses), and others such as \ion{O}{1}
(gold stars) and the combination of limits (black diamonds). 
}
\label{fig:metallit}
\end{figure}

In Figure \ref{fig:metallit}, we plot the compilation of 195 metallicities literature abundance values as a function of redshift.
The figure outlines which elements are used to obtain [M/H] and the uncertainties of those measurements. 
The symbols and colors indicate the origin of the [M/H] measurements, with blue triangles representing \ion{S}{2}, 
green filled circles representing \ion{Si}{2}, red plus crosses representing \ion{Fe}{2}, brown boxes representing \ion{Zn}{2}, 
and gold stars representing \ion{O}{1}, and combination of limits (black diamonds). 
We also present these values in Table \ref{metal:tab3}, where 
column (1) gives the quasar name, 
column (2) gives the DLA absorption redshifts, and 
column (3) gives the logarithm of the measured {\nh} values.
Columns (4) and (5) give the alpha element flag and metallicity, 
and column (6) and (7) give the Fe flag and metallicity. 
Column (8) and (9) give the flag and final metallicity used in this paper, 
and column (10) cites the references. 
Similar to Table \ref{metal:tab3}, the flags in columns (4) and (6) state which transitions are used and if they are detections or limits,
as described in the Table notes. The uncertainties in the table include the uncertainty in {\nh} added in quadrature. 

\input{tab3.tex}

\begin{figure}
\center{
\includegraphics[scale=0.45, viewport=15 5 495 360,clip]{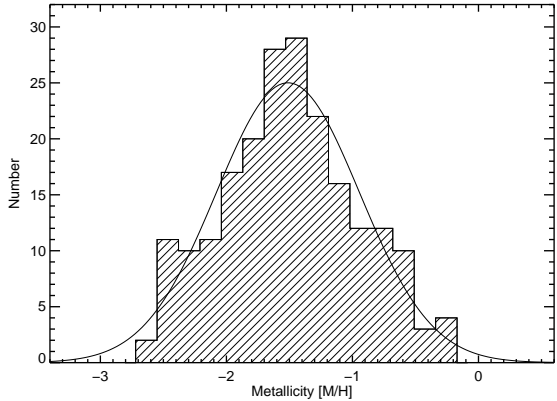}
}
\caption{Metallicity distribution function for 207 DLA metal abundances, [M/H], including abundances from both the literature and our new observations. 
Only DLAs with $z>1.5$ are used since $z<1.5$ DLAs are metallicity biased.
The black line is a Gaussian fit to the data, with a mean [M/H] of -1.51 and a width $\sigma=0.57$. 
Note the very low probability of detecting a system at [M/H]$<-3$ if this Gaussian is a proper description of the gas metallicity at low values.
}
\label{fig:metalhist}
\end{figure}

\begin{figure}
\center{
\includegraphics[scale=0.45, viewport=15 5 495 360,clip]{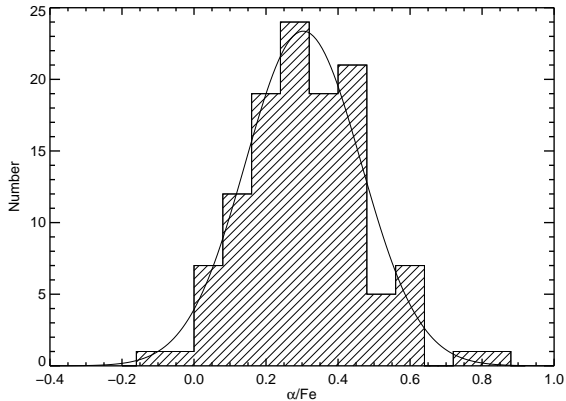}
}
\caption{Alpha enhancement, $\alpha$/Fe, distribution function of 154 DLAs, including abundances from both the literature and our new observations. 
Only DLAs with $z>1.5$ and [M/H]$<-1$ are used since $z<1.5$ DLAs are metallicity biased, and Fe measurements with [M/H]$>-1$ are depleted by dust.
The black line is a Gaussian fit to the data, with a mean $\alpha$/Fe of 0.30 and a width $\sigma=0.16$.
\vspace{-3mm}
}
\label{fig:alphahist}
\end{figure}

We combine the 195 literature abundance values with our new ESI and HIRES observations for a total of 241 abundances. We investigate the 
abundance distribution of DLAs, and here only include the 207 DLAs with $z>1.5$, as $z<1.5$ DLAs
are selected for the presence of strong metal lines and are therefore biased. 
In Figure \ref{fig:metalhist}, we plot the metallicity distribution function for DLAs, and the black line is a Gaussian fit with a mean [M/H] of -1.51 and a width $\sigma=0.57$. 
We note that this dispersion is not due to the uncertainties in the measurements, with a mean uncertainty of 0.13$\pm$0.05. 
 In addition, we consider the distribution of $\alpha/$Fe values for DLAs with $z>1.5$ that have an abundance of both an alpha and an iron peak element. 
 For these DLAs, we also require that $[\alpha/{\rm H}]<-1$, as 
Fe is depleted by dust in DLAs with higher metallicities \citep{Vladilo:2011}. 
Figure \ref{fig:alphahist} plots the [$\alpha/$Fe] distribution function for 154 DLAs, and the 
 black line is a Gaussian fit with a mean [$\alpha/$Fe] of 0.30 and a width $\sigma=0.16$. 
The mean uncertainty of [$\alpha/$Fe] is 0.08$\pm$0.05, which is smaller than the dispersion of [$\alpha/$Fe],
suggesting there may be intrinsic variations between DLAs. Nonetheless, we also calculate the standard deviation of the mean, and 
find that the mean value of [$\alpha/$Fe]$=$0.30$\pm$0.02 (see also \S\ref{comp}).

\begin{figure*}
\center{
{\includegraphics[scale=0.48, viewport=15 5 495 360,clip]{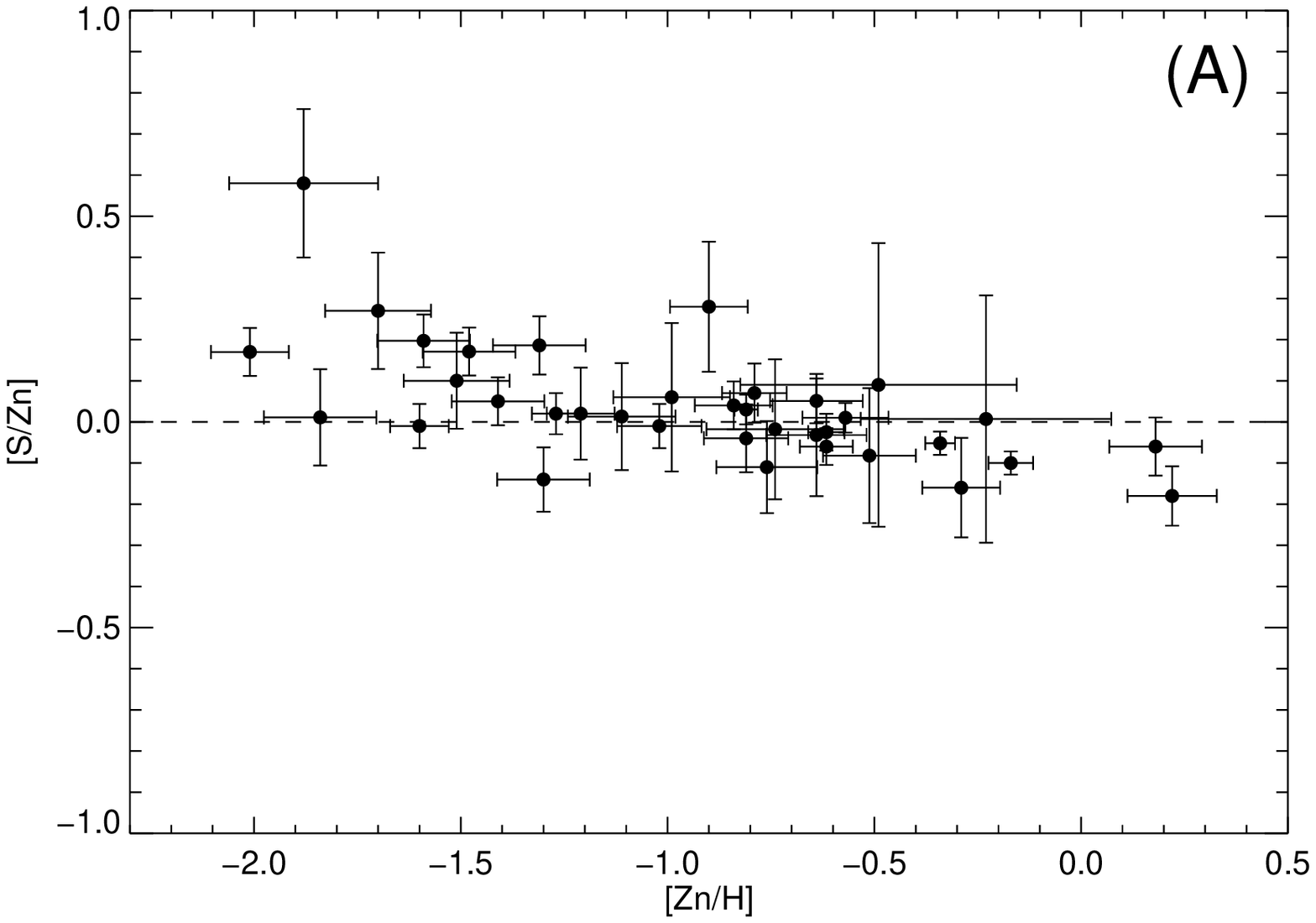}}
\hfill
{\includegraphics[scale=0.48, viewport=15 5 495 360,clip]{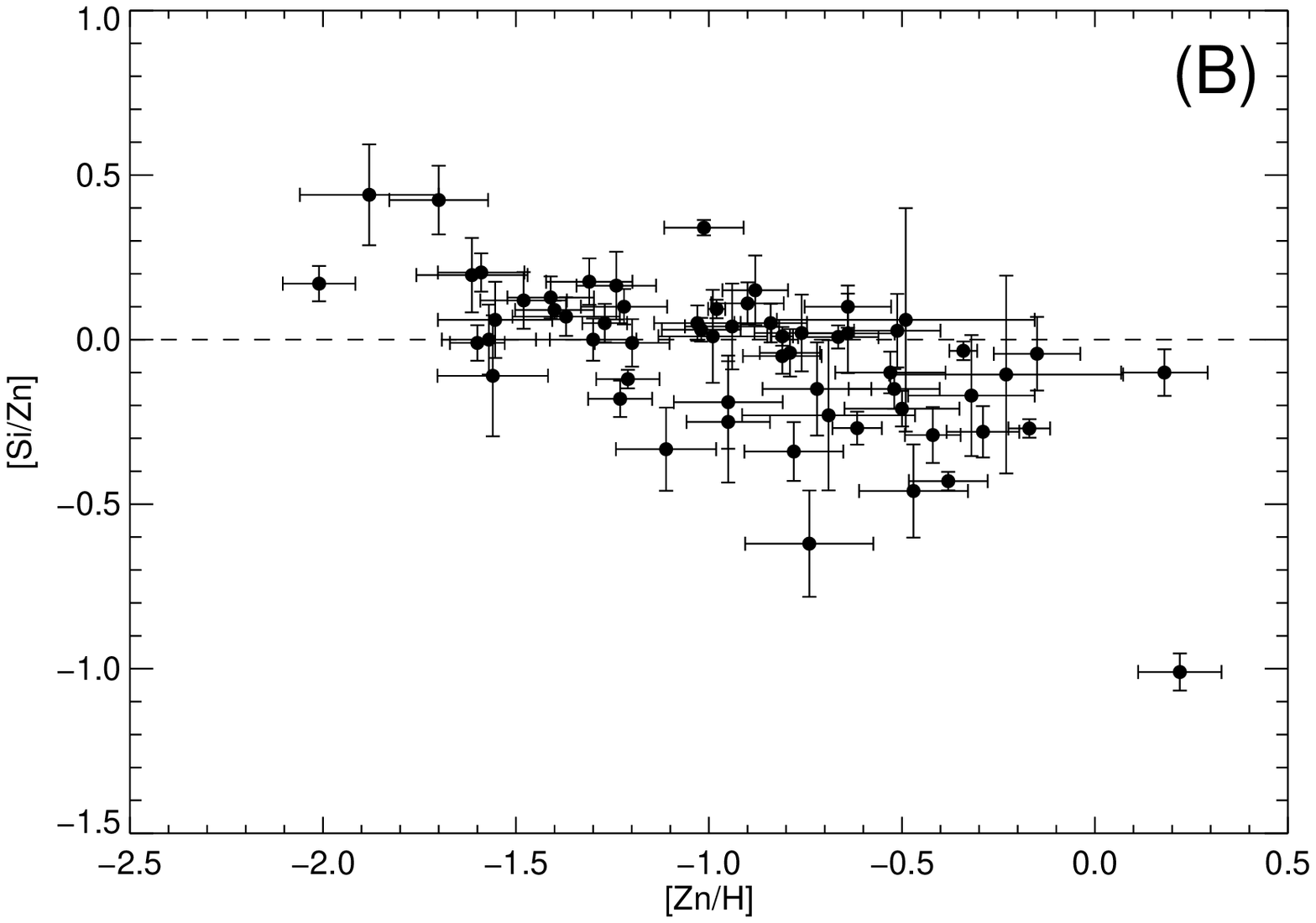}}
\hfill
{\includegraphics[scale=0.48, viewport=15 5 495 360,clip]{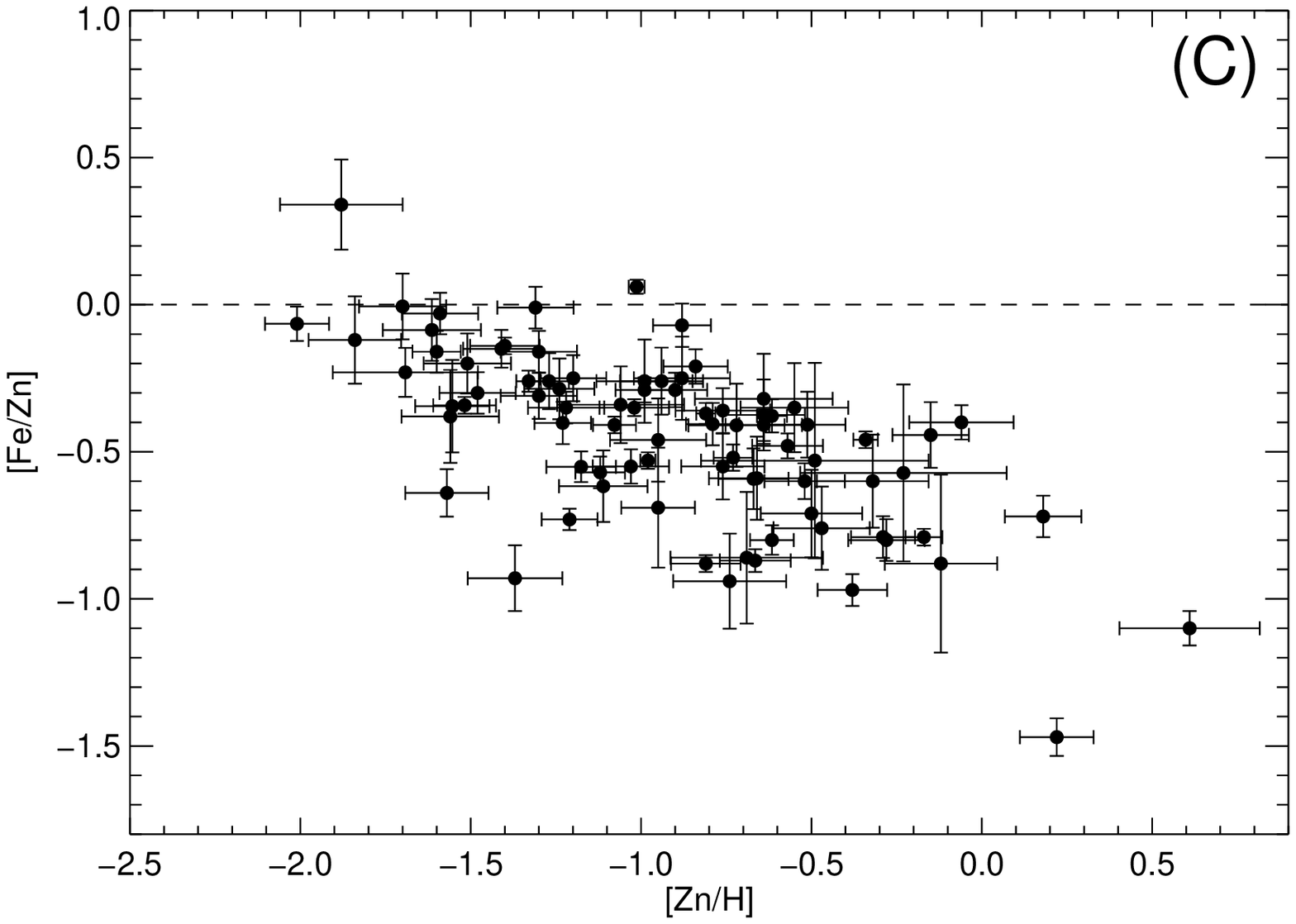}}
\hfill
{\includegraphics[scale=0.48, viewport=15 5 495 360,clip]{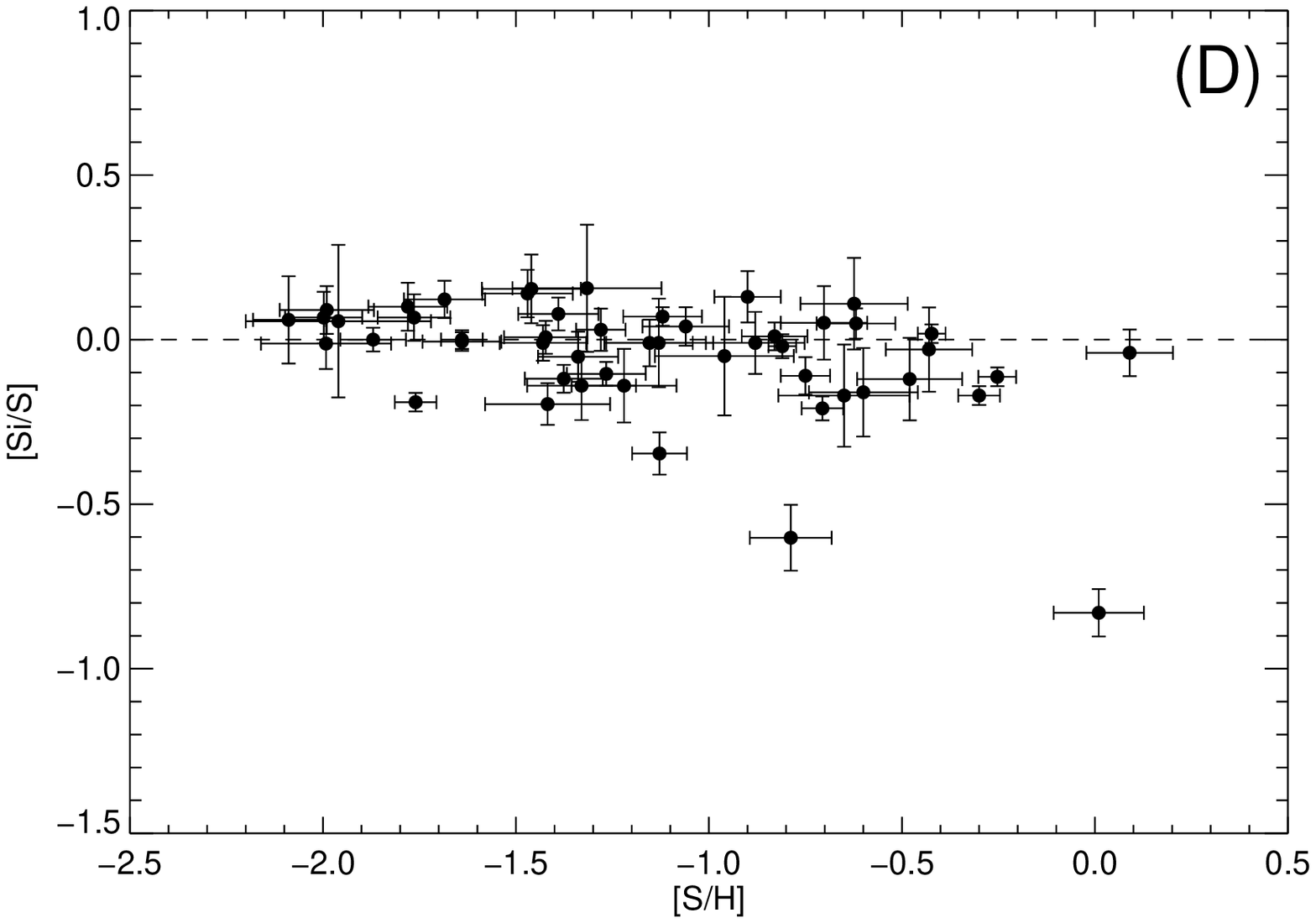}}
} 
\caption{Comparison of different elements used to determine metallicities of DLAs. 
Panel (a) compares S and Zn, Panel (b) compares Si and Zn, Panel (c) compares Fe and Zn, and Panel (d) compares
Si and S. The Kendall Tau test shows that they are all technically correlated. The correlation in Panel (a) may be spurious,
and that of in Panel (b) may be due to selection effects. Panel (c) is clearly correlated, although at low metallicities flattens out.
Panel (d) suggests that Si is not more significantly depleted than S,
aside from a few special cases. }
\label{fig:comp}
\end{figure*}

\subsection{Comparisons of Elemental Abundances}
\label{comp}

We use a number of different ions for determining our elemental abundances in the order O, S, Si, Zn, and Fe as described in \S\ref{order}. 
In this subsection we justify our practice of mixing these elements in
our analysis of DLA metallicities by comparing S, Si, and Fe to Zn in Figure \ref{fig:comp}.
These measurements are based on a different set of DLAs than those described in \S\ref{lit}. These measurements are based
only on very high resolution measurements from HIRES or UVES as compiled by \citet{Vladilo:2011}, with additional data from the present study and from \citet{Wolfe:2008}. 
We note that for these comparisons we are not concerned about any metallicity biases, and therefore use all data that is available.

In Figure \ref{fig:comp} we compare different elemental abundances, and any any observed trends may be related to dust, nucleosynthesis, and even ionization. 
Panel (a) in Figure \ref{fig:comp} plots [S/Zn] versus [Zn/H], Panel (b) plots [Si/Zn] versus [Zn/H], Panel (c) plots [Fe/Zn] versus [Zn/H], and Panel (d) plots
[Si/S] versus [Si/H]. A standard Kendall Tau test shows that they are all correlated, with the strongest anti-correlation in Panel (c) for [Fe/Zn] and [Zn/H]. 
However, we must correct the correlation coefficients for the fact that the ratios comprising the x and y axes contain the same elements.
For instance, Panel (a) in Figure \ref{fig:comp}
contains Zn in both axes, and therefore the Kendall Tau test should find a correlation even if there is none. To account for this effect,
we calculate partial correlation coefficients following the prescription in \citet{Akritas:1996}, and find that only  [Si/Zn] versus [Zn/H] and [Fe/Zn] versus [Zn/H] (Panels (b) and (c))
are anti-correlated at greater than 3$\sigma$. However, we caution the reader that for all but [Fe/Zn] versus [Zn/H], we are looking for trends close to the limits
of what the data can provide. While we correct for the partial correlation,  a small number of outliers can effect the correlation. 

First, we find a minor  2-3$\sigma$ anti-correlation in Panel (a) for [S/Zn] and [Zn/H]. While this is not very significant, there does appear to be a set of values above the dashed
line at low metallicity, and below the dashed line at high metallicity. This may be due to a nucleosynthetic effect, or due to small number statistics. 
Panel (b) shows what appears to be an anti-correlation between [Si/Zn] and [Zn/H], and to investigate this
we therefore also compare Si with S in panel (d). Panel (d) shows a 2$\sigma$ anti-correlation between [Si/S] and [S/H], although this anti-correlation 
is sufficiently weak and near zero that there may be no trend. In fact, other than a few outliers, there is no obvious correlation visible by eye. 
This suggests that the minor anti-correlation observed in Panel (b) may be a selection 
effect, or some currently not understood effect. We therefore conclude that if there is any dust depletion in Si like in the local ISM, it is a small effect, otherwise it would 
have been clearly observed in panel (d). We note that it is of course possible that any individual DLA is depleted, which may explain the two outliers in the figure. 
Lastly,  [Fe/Zn] versus [Zn/H] are anti-correlated as expected because Fe is depleted, and is
expected to show higher depletion at higher metallicities. 
In fact, at low metallicities, Panel (c) shows that there is a 
flattening of the [Fe/Zn] trend, suggesting that depletion in Fe is not large for low metallicities, 
as found previously by \citet{Vladilo:2011}.  We therefore use S, Si, Zn, and Fe to measure elemental abundances, although only use Fe at lower metallicities. 

When using different elements for determining metallicities, a correction must be applied when combining 
$\alpha$-enhanced elements with Fe-peak elements (see \S4.4). While S and Si are $\alpha$-enhanced in the Milky Way, 
the nature of Zn is less clear. 
The comparisons in Figure \ref{fig:comp} shows that [Fe/Zn] is offset from 0 even at low metallicites, while [Si/Zn] and [S/Zn] are around 0, suggesting
that Zn is enhanced in a way similar to $\alpha$-elements, rather than Fe-peak elements. This supports treating Zn as an  $\alpha$-element in our metallicity 
evolution plots, as well as in our measurement of the  $\alpha$-enhancement of DLAs. 

\begin{figure}
\center{
\includegraphics[scale=0.45, viewport=15 5 495 360,clip]{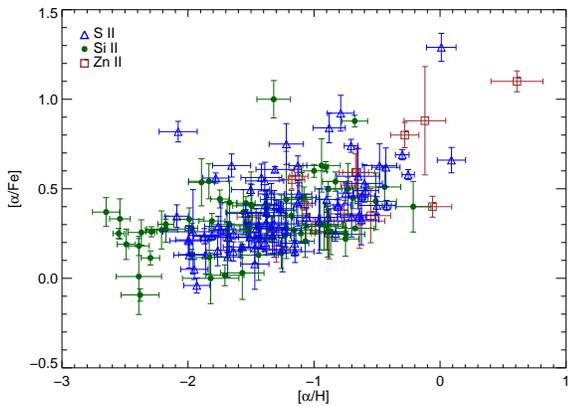}
}
\caption{$\alpha$-enhancement, [$\alpha$/Fe], as a function of the metallicity using Si, S, and Zn as alpha elements.
At low metallicities ([$\alpha$/H] $<-1.0$) the [$\alpha$/Fe] distribution flattens out to a mean value of $0.27\pm0.02$.
At higher metallicities, there is a correlation of [$\alpha$/Fe] with [$\alpha$/H], which may be caused by the depletion of Fe.
This systematic offset of [$\alpha$/Fe]  from 0 in the regime where Fe is not depleted (low metallicity) suggests that DLAs are alpha enhanced. 
}
\label{fig:alphafe}
\end{figure}

We can also use this same high resolution dataset as in Figure \ref{fig:comp} to explore the $\alpha$-enhancement of DLAs with higher precision than in Figure \ref{fig:alphahist}. 
Figure \ref{fig:alphafe} plots the $\alpha$-enhancement, [$\alpha$/Fe] as a function of the metallicity using Si, S, and Zn as alpha elements. 
This shows that while there is a large scatter in the [$\alpha$/Fe] values, at low metallicities ([$\alpha$/H] $<-1.0$) the [$\alpha$/Fe] distribution flattens out to a mean value of $0.27\pm0.02$,
where the uncertainty is the standard deviation of the mean. A Gaussian fit to [$\alpha$/Fe] has a mean of [$\alpha$/Fe] of 0.26 and a width $\sigma=0.12$, which
matches that found for the larger lower resolution unbiased sample used in most of this paper. 
At higher metallicities, there is a correlation of [$\alpha$/Fe] with [$\alpha$/H], possibly caused by the depletion of Fe, which is why we limit ourselves to lower metallicity DLAs ([M/H] $<-1$)
when comparing the $\alpha$-enhancement of DLAs (see \S\ref{halo}). This systematic offset of [$\alpha$/Fe]  from 0 in the regime where Fe is not depleted suggests that DLAs are alpha enhanced. 

\subsection{Evolution of Cosmic Metallicity}
\label{cosmic}

\begin{figure*}
\center{
\includegraphics[scale=0.75, viewport=15 5 495 360,clip]{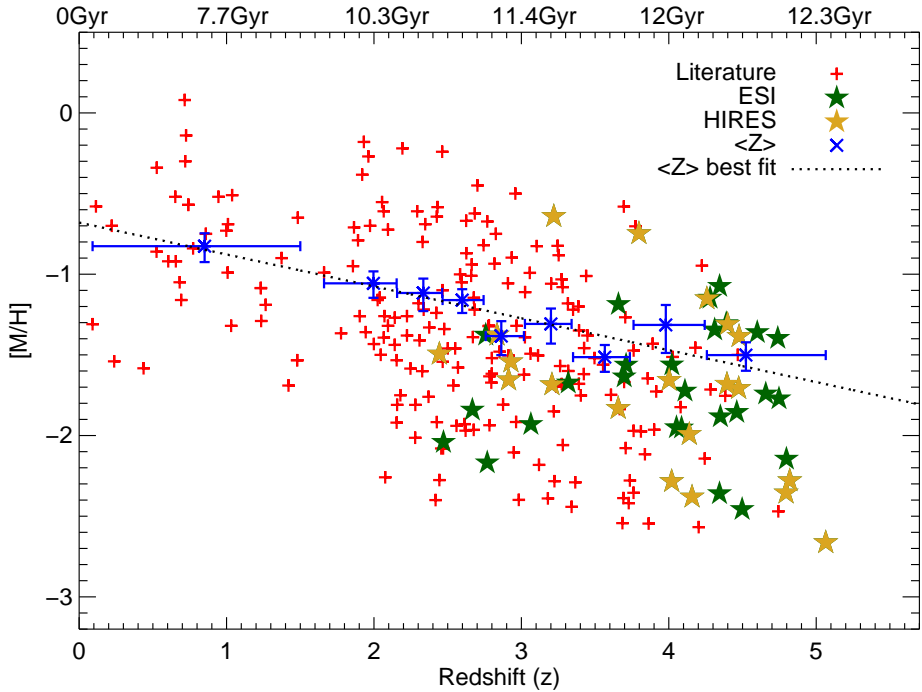}
}
\caption{DLA metal abundance, [M/H],  versus
redshift. Red plus signs from the literature. 
The metallicities from ESI are green stars and those from HIRES are gold stars.
The 9 blue crosses with error bars show cosmic metallicity,
{\Z}, where horizontal error bars are determined such that there are equal numbers
of data points per redshift bin. The vertical error bars 
represent 1$\sigma$ confidence levels given from our
bootstrap analysis. 
Black straight line is linear fit to the {\Z} data points.
}
\label{fig:metal}
\end{figure*}

We trace the build-up of heavy elements in neutral gas across cosmic time by investigating the metallicity evolution of DLAs out to $z\approx5$. 
The following analysis presents the results under the assumption and expectation that dust obscuration has a modest effect on the majority of
DLA sightlines \citep{Ellison:2001a, HerbertFort:2006, Jorgenson:2006, Meiring:2006, Khare:2007}.
In Figure \ref{fig:metal}, we plot the metallicity as a function of redshift for both our measurements and those from the literature with a total of 241 abundances.
Specifically, we plot the metallicities from the literature, and overplot our new [M/H] measurements 
as stars, with ESI measurements as green stars and HIRES measurements as gold stars. Of these new [M/H] measurements, 
30 are at $z>4$. 

There are three interesting features in this plot:
1) there exists a large dispersion in [M/H] at all redshifts, 
2) there appears to be a metallicity ``floor'' at [M/H]$\approx-3$, and 
3) the average [M/H] decreases with increasing redshift. 
The large dispersion in Figure \ref{fig:metal}  is not an observational error, but is due to intrinsic scatter amongst the DLAs. 
The dispersion in [M/H] is $\sim0.5$ dex, and does not appear to evolve with redshift. 
The metallicity ``floor'' occurs at [M/H]$\approx-3$,
with no DLAs existing at [M/H] $<$ $-$2.8 despite the sensitivity of
our spectra for finding DLAs with [M/H] lower than $-4$.
The implications of the scatter and the metallicity``floor'' will be discussed in \S\ref{discuss_metalevol}. 

The evolution of [M/H] is investigated by computing the cosmological mean metallicity, {\Z}, where
\begin{equation}
\langle {\rm Z}\rangle= \log \left( \sum_i 10^{[M/H]_i} N({\rm H I})_i / \sum_i N({\rm H I})_i \right), 
\end{equation}
 
\noindent and $i$ represents each bin of DLAs as a function of redshift \citep{Lanzetta:1995, Prochaska:2003b}.
Stated differently,  $\langle {\rm Z}\rangle=  \log(\Omega_M/\Omega_{HI})-\log(\Omega_M/\Omega_{HI})_\odot$,
where $\Omega_M$ and $\Omega_{HI}$ are the comoving densities of metals and of atomic hydrogen. 
We bin the data such that each bin contains an equal number of DLAs (26 DLAs/bin), 
except for the lowest and highest redshift bins which contain 34 and 25 DLAs respectively. 
We maximize the number of bins while keeping the resultant values uncorrelated and require a sufficient number of DLAs per bin such that 
we obtain Gaussian distributions in the bootstrap analysis for determining our uncertainties. 
{\Z} is plotted as blue crosses in Figure \ref{fig:metal},  and the error bars represent 1$\sigma$ confidence limits based on a 
bootstrap error analysis described below. 

The {\Z} statistic is dominated by the systems with the largest {\nh} and [M/H] values (although we find no correlation of  {\nh} with [M/H]),
and therefore the measurement uncertainty is determined by sample variance rather than statistical error. 
We calculate the uncertainties on {\Z} by using a bootstrap method, where we replace a random fraction 
($1/e \approx 37\%$) 
of the DLAs in each bin with other DLAs from that bin. We repeat this 1000 times, and then take the 
standard deviation of all the column density weighted metallicity measurements. 
We chose the number 1000 because it is sufficiently large such that the distribution of {\Z} resembles a Gaussian distribution, and increasing this number does not change the result.
This method is conservative, and includes all the uncertainties associated with the variance of the sample, if the distribution of 
the bootstrap values are Gaussian.  
We note that we have sufficiently high {\nh} DLAs in our sample such that the distribution is close to Gaussian, and 
the results are insensitive to the addition of any single DLA \citep[see also][]{Prochaska:2003b}. The only exception to this
is the inclusion of J1201+2117 at $z_{\rm abs}=3.79$, which has an extreme velocity width showing absorption lines over a 1500 km s$^{-1}$ range. 
Regardless of how we bin our data, the cosmic metallicity bin including this DLA is slightly elevated as evidenced in the $z\sim4$ bin in Figure \ref{fig:metal}.
The effect of this DLA on the binned cosmic metallicity is included in the bootstrap analysis, resulting in a slightly larger uncertainty for this value. 

The black dotted line in Figure \ref{fig:metal} is a linear fit to the {\Z} values and their bootstrap uncertainties. 
In order to properly treat the uncertainties, we perform a chi-square minimization fit to a  logarithmic function on the linear values of the cosmic metallicities 
and their associated uncertainties from the bootstrap. We find that  $\langle {\rm Z}\rangle=(-0.22 {\pm} 0.03)z-(0.65 {\pm}0.09)$, excluding rounding
errors. This is a reasonable, although non-physical, fit to the data to describe the slope of the evolution. 
This yields an $6.7\sigma$ detection in the evolution of {\Z}, which is a significant improvement
over the previous $\sim3\sigma$ detections by \citet{Prochaska:2003b}, \citet{Kulkarni:2005}, \citet{Kulkarni:2007}, and \citet{Kulkarni:2010}. 
While the values of the fit are extremely similar to what was found previously by \citet{Kulkarni:2010}, the significance of the evolution is improved 
upon by the larger lever arm obtained by sampling DLAs at higher redshift, more DLAs at lower redshift, and a larger sample size. We note that 
the sample compiled from the literature (see Table \ref{metal:tab3}) is somewhat different than that compiled by \citet{Kulkarni:2010}, 
since we exclude some studies due to potential metallicity bias, which may also affect the results.  Regardless, there is good agreement of the slope 
and intercept of these semi-independent studies.

The larger the number of cosmic metallicity measurements, the better we can determine the evolution of the cosmic metallicity. However, if we reduce the number of DLAs per bin too significantly, then our binned values will be plagued by small number statistics, and the bootstrap may not result in Gaussian distributions. As a test to investigate our sensitivity to binning, we apply a running bin with 30 DLAs per bin. Each bin adds and removes one DLA while increasing in redshift, resulting in 202 bins. This methodology results in correlated values, as each DLA is used many times, and thus is not used as our primary technique. However, it is a good test of the effect of binning on our data. This yields $\langle {\rm Z}\rangle=(-0.19)z-(0.72)$, where we omit the error bars due to the heavy correlation of the data. The slope and intercept are very similar to the above values, and are well within the uncertainties of the uncorrelated data. This suggests that binning is not strongly affecting our results.  

In the fit of the cosmic metallicity, we make no corrections for the potential metallicity bias in our lowest redshift bin mentioned in \S\ref{lit}. 
Specifically, the DLAs at $z\lesssim1.5$ are generally selected based on their strong \ion{Mg}{2}, rather than on
the damped {\lya} line. The two selection methods are not equivalent, and DLAs selected for strong \ion{Mg}{2} absorption systems are generally
more metal rich than {\lya} selected systems \citep{Nestor:2003, Nestor:2008, Murphy:2007}. Correcting for this bias of 
$\sim0.1$ dex \citep{Nestor:2008} results in $\langle {\rm Z}\rangle=(-0.19 {\pm} 0.03)z-(0.74 {\pm}0.09)$, which is a 
$6.3\sigma$ detection in the evolution of {\Z}. 
On the other hand, removing the lowest redshift bin completely, and thereby only using DLAs with $z>1.5$, results in 
$\langle {\rm Z}\rangle=(-0.22 {\pm} 0.05)z-(0.66 {\pm}0.15)$, which is a $4.6\sigma$ detection in the evolution of {\Z}.
We note that all the fit values are completely consistent with the previous measurements by \citet{Prochaska:2003b},
confirming that the mean metallicity of the universe in neutral gas is doubling about every billion years at $z\sim3$.

\subsection{Comparison of  Metallicities of DLAs and Halo Stars}
\label{halo}

DLAs are widely believed to be the progenitors of today's disk galaxies and 
act as neutral gas reservoirs for star formation at high redshifts 
\citep[][]{Nagamine:2004b, Nagamine:2004a, Wolfe:2006, Wolfe:2008, Rafelski:2011}.
Stars forming {\it in situ} out of low metallicity DLA gas would result in low metallicity stars, which should be observable in the Milky Way today. 
However, previous studies found that the metallicity distribution of DLAs disagree with known stellar populations in the 
Galaxy \citep{Pettini:1997, Pettini:2004, Pettini:2006}. On the other hand, recent studies of metal-poor DLAs suggests that their [C/O] distribution 
is in good agreement with metal-poor halo stars \citep{Penprase:2010,Cooke:2011}.

Here we investigate whether the increase in sample size and redshift yields a metallicity distribution 
in better agreement with Galactic stellar populations. We compare the metallicity distributions of DLAs with the thin disk stars, thick disk stars, and halo stars,
although we mainly focus on the halo stars as the metallicities of the other two are systematically too high (as we show below). 
We limit ourselves to DLAs at $z>2$, corresponding to an age of $\gtrsim$10 billion years, similar to the ages of halo stars in the Milky Way. 
However, the results are insensitive to including DLAs at $z>1.5$. 

Halo stars generally have low metallicities and are often selected based on 
their metallicities \citep[e.g.][]{Beers:1995}. However, there is an overlap in the metallicity of different stellar components 
in the galaxy \citep{Unavane:1996, Chiba:2000}. 
Therefore, \citet{Venn:2004} select stars based purely on their kinematics, compile a large dataset from a number
of publications, and provide probabilities of each star consisting of a thin disk, thick disk, or halo star, based on their kinematics. 
Here, we select stars from the \citet{Venn:2004} sample of local stars with an 80\% or better probability of being identified as
each type of star and having both $\alpha$ and Fe element abundances measured.
This results in a sample of 201 thin disk stars, 109 thick disk stars, and 207 halo stars
\citep{Venn:2004, Edvardsson:1993, Hanson:1998, Fulbright:2000, Fulbright:2002, Reddy:2003}. 

While \citet{Venn:2004} sample does not select stars based on their metallicity, the sample is still metallicity biased
as the studies it uses preferentially target metal poor objects. This is a well known bias, and its effects are discussed in depth for
a different sample of halo stars by \citep{Schorck:2009}. In that case, the metallicity distribution function was offset by 0.3 dex lower than the intrinsic distribution. 
While larger less biased samples of stellar metallicities and classifications exist, these data have not been made public \citep[e.g.][]{Carollo:2010}. 
In addition, while our DLA sample is not
metallicity biased per redshift bin, it is slightly biased as a whole due to
the metallicity evolution with redshift. If we had targeted additional high
or low redshift DLAs, it would slightly change our metallicity distribution function by decreasing or increasing the mean metallicity respectively. 
Since we are relatively evenly sampled at $z>2$, we expect this to be a small effect. 
Moreover, a comparison of stellar metallicities with those of DLAs should be approached with caution, as all the stars from any given 
stellar population in the Milky Way could have formed out of just a small number of DLAs. Also, while it is logical to assume that 
the Milky Way is a representative object, there will be variations between galaxies that affect the comparison. Hence, 
while we expect the metallicities of stars formed out of DLA gas to be similar to those of stars in the Milky Way, they do not need to be 
exact matches. It is, however, illustrative to compare the metallicity of DLAs with those of different stellar populations.  

\begin{figure}[h!]
\center{
\subfigure[Thin Disk Stars]{\includegraphics[scale=0.48, viewport=15 5 495 360,clip]{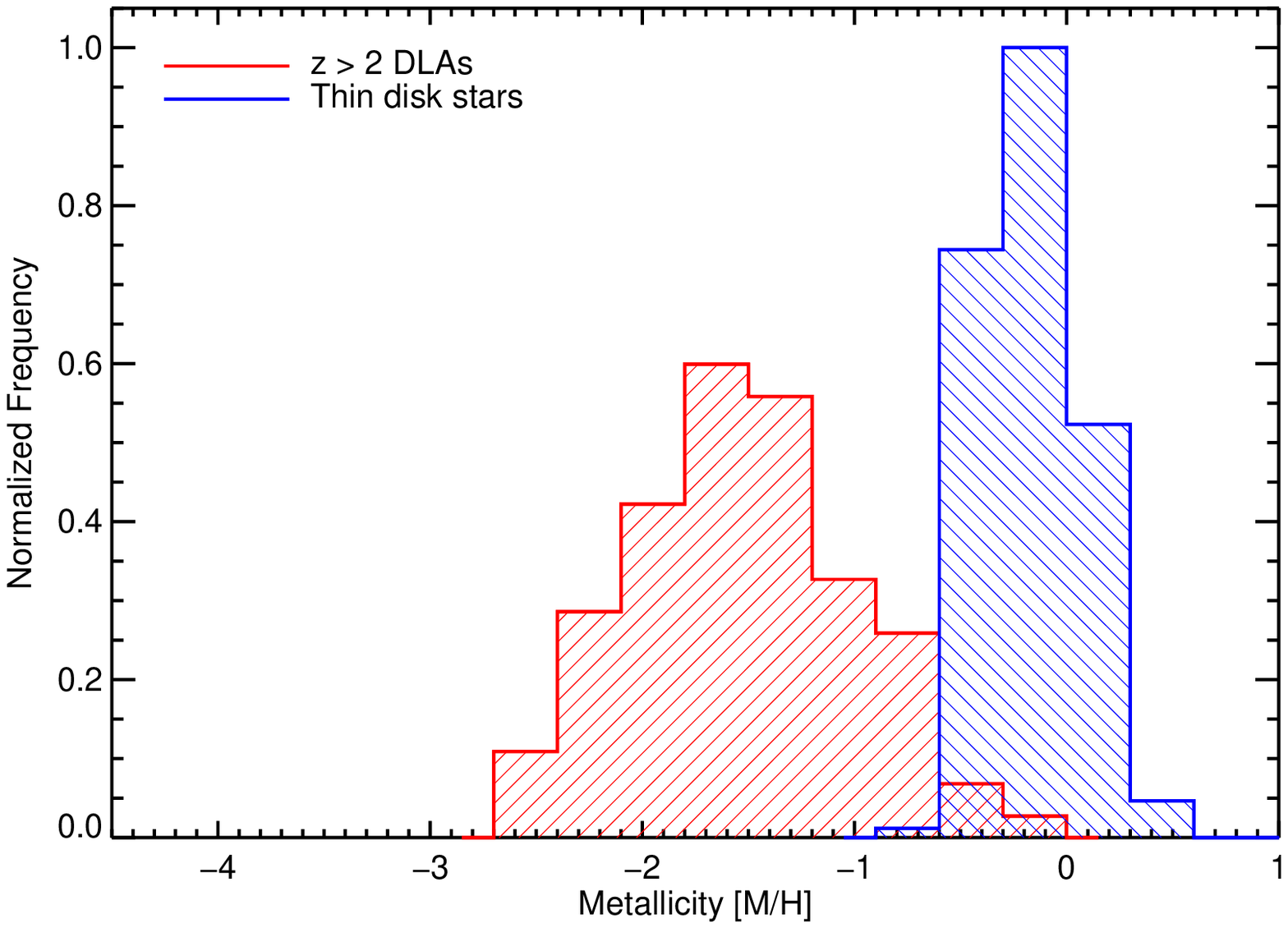}}
\hfill
\subfigure[Thick Disk Stars]{\includegraphics[scale=0.48, viewport=15 5 495 360,clip]{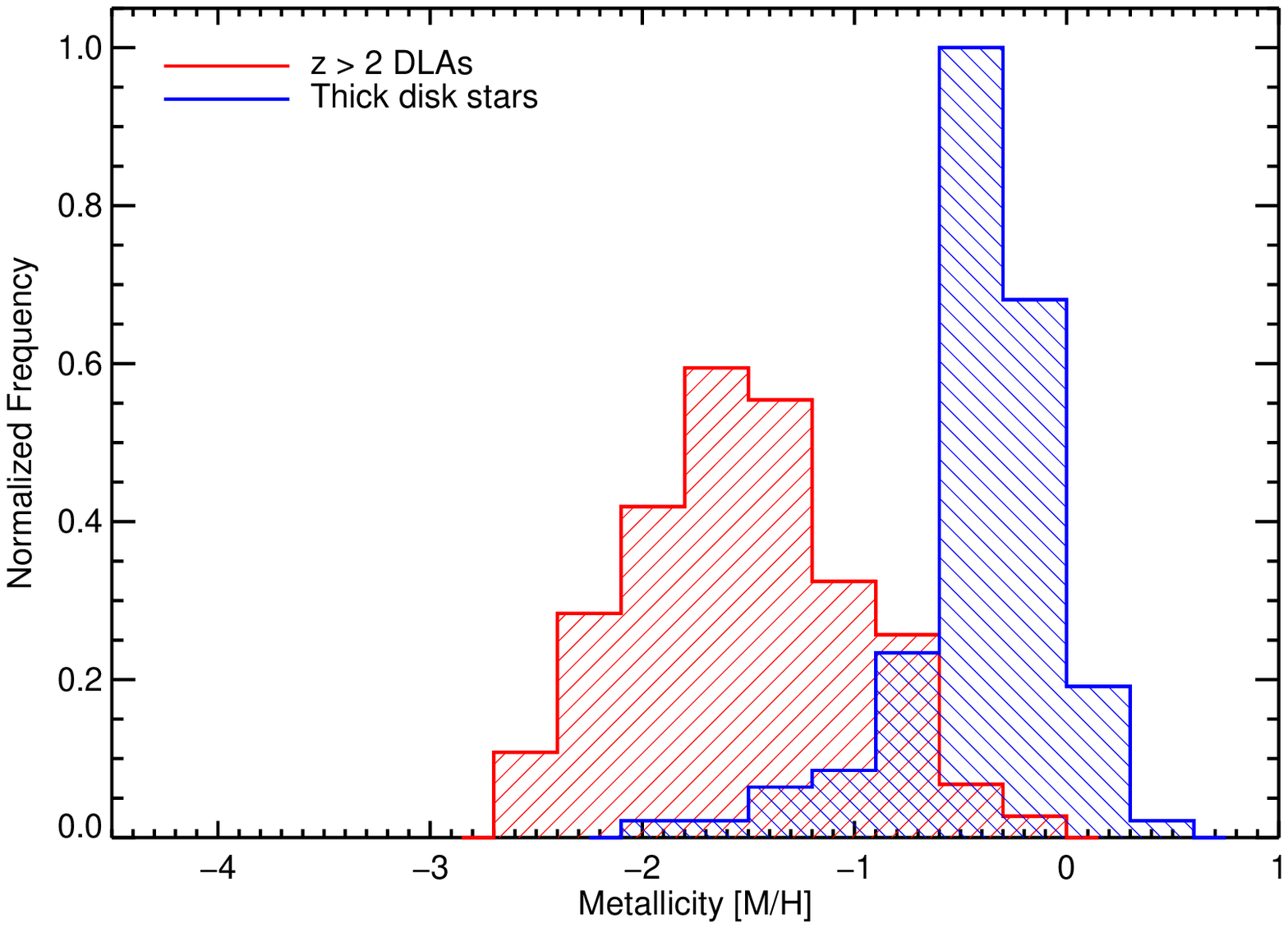}}
\hfill
\subfigure[Halo Stars]{\includegraphics[scale=0.48, viewport=15 5 495 360,clip]{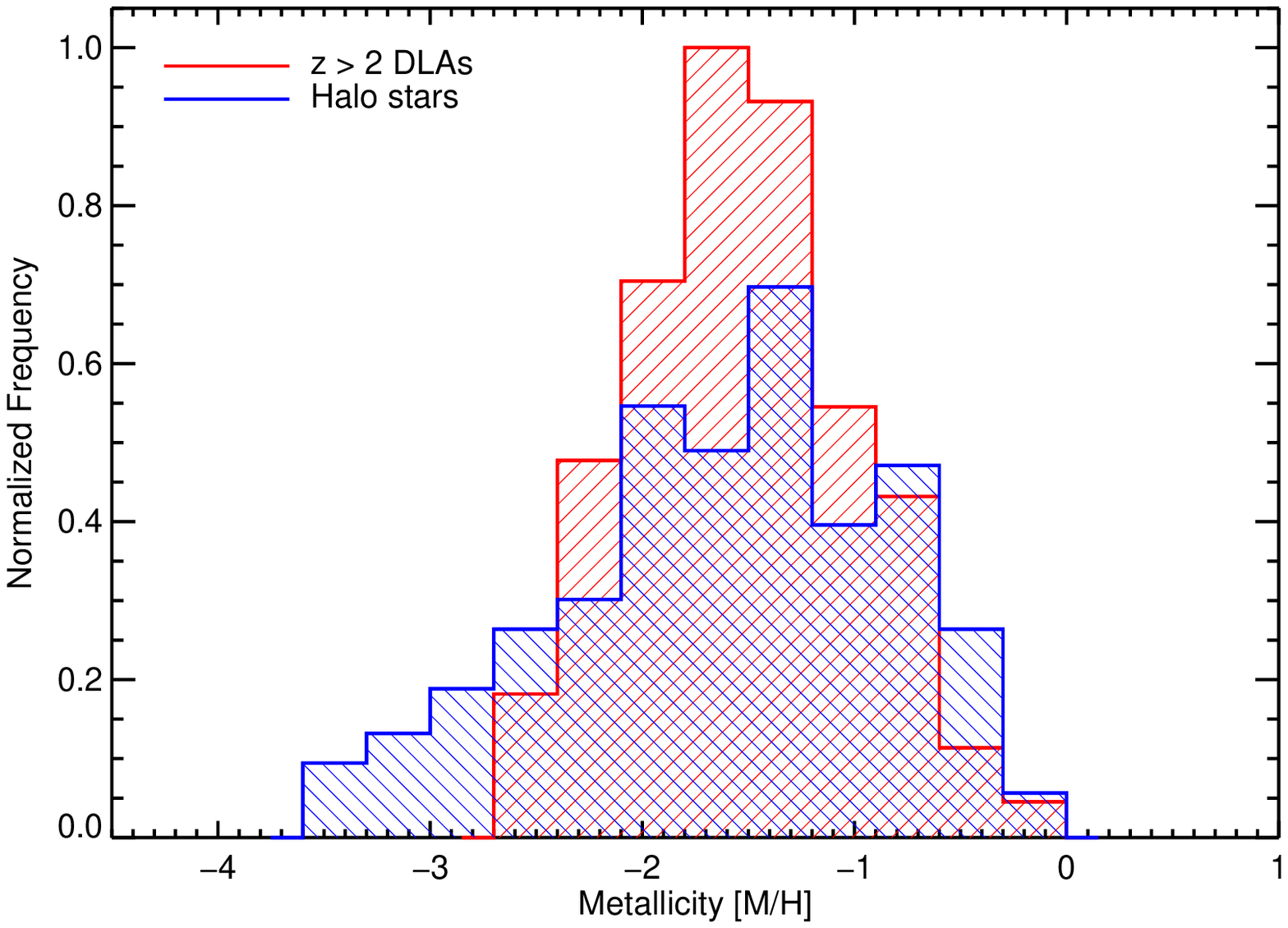}}
} 
\caption{Histograms comparing abundances of 161 DLAs at $z>2$ with a) 201 thin disk stars,  
b) 109 thick disk stars,  and c) 207 halo stars \citep{Venn:2004}. 
The histograms are normalized to have equal area and a maximum value of 1.0
The red left shaded region represent the DLAs and the blue right shaded region represent the thin disk, thick disk, and halo stars. 
The only population of stars in the Milk Way that is consistent with being drawn
from the same parent population are the halo stars.
}
\label{fig:halo}
\end{figure}

In Figure \ref{fig:halo}, we plot histograms comparing the abundances of 195 DLAs at redshifts $z>2$ (red), 
with a) thin disk stars, b) thick disk stars, and c) halo stars (blue). For the DLAs, we use the same metal abundances 
as used in Table \ref{metal:tab2} and Figure \ref{fig:metal}. For the stars, we use the $\alpha$-element abundances
tabulated in \citet{Venn:2004}. The histograms are normalized to have equal area and a maximum value of 1.0. 
We compare the stars to DLAs with $z>2$, as the majority of the halo stars are believed to have formed by 
then \citep{Freeman:2002, Bullock:2005, Robertson:2005, Johnston:2008}. 
This cut includes the vast majority of our sample, and therefore has a negligible effect on the comparisons. 
A visual comparison of histograms immediately rules out the possibility of any agreement of the metallicity distributions
of the thin disk or thick disk stars with DLAs. However, the halo star metallicity distributions look similar, and the 
median [M/H] of the two distributions are basically the same, with values of -1.54 and -1.53 for the DLAs and halo stars respectively.  

We apply the Kolmogorov-Smirnov (K-S) test to find the probability of each of the distributions being drawn 
from the same parent population as the DLA metallicities. Similar to the visual comparison, the thin disk and thick disk stars are ruled out
at $\sim0$\% probability, i.e., the null hypothesis can be rejected at a high confidence level.
However, the halo stars have a probability of 3\%, i.e., the null hypothesis cannot be rejected with more 
than 2$\sigma$ confidence. The two samples are therefore consistent with being drawn
from the same parent population.
Furthermore, we reiterate the warning mentioned above, 
that the two distributions do not need to match exactly. For instance, the metal poor tail shown here for halo stars is 
not evident in the metallicity distribution function from SDSS \citep[see Figure 2 in][]{Carollo:2010}, and is due to the 
metal poor bias mentioned previously. 
In any case, this comparison hints at an overlap between the 
metallicity distributions of high redshift DLAs and halo stars in the Milky Way. We discuss the implications of this in \S\ref{halocomp}.

Another possible comparison of DLAs and halo stars can be made by studying the chemistry of the two populations. 
Specifically we can compare the ratio of $\alpha$-elements to Fe abundances to determine whether gas is $\alpha$-enhanced. 
$\alpha$-elements are produced in high mass stars and then are ejected by Type~II supernovae (SNe),
while Fe is produced in both Type~II and Type~Ia SNe. 
Therefore, if stars are formed shortly after the Type~II SNe explode,
but before the Type~Ia SNe do, they will have enhanced [$\alpha$/Fe] ratios \citep{Matteucci:2001}. 
While the exact lifetimes of SNe is an active area of research, Type~II SNe (and other $\alpha$-element producers such as Type Ib and Ic SNe) generally need 10$^6$-10$^7$ years to explode, while Type Ia SNe need more than 10$^8$-10$^9$ years \citep{Kobayashi:2009}. What is important is that Type Ia SNe take significantly more time to produce the Fe elements than the other SNe require to produce $\alpha$-elements.

\begin{figure}
\center{
\subfigure[$z>2$ DLAs]{\includegraphics[scale=0.48, viewport=15 5 495 360,clip]{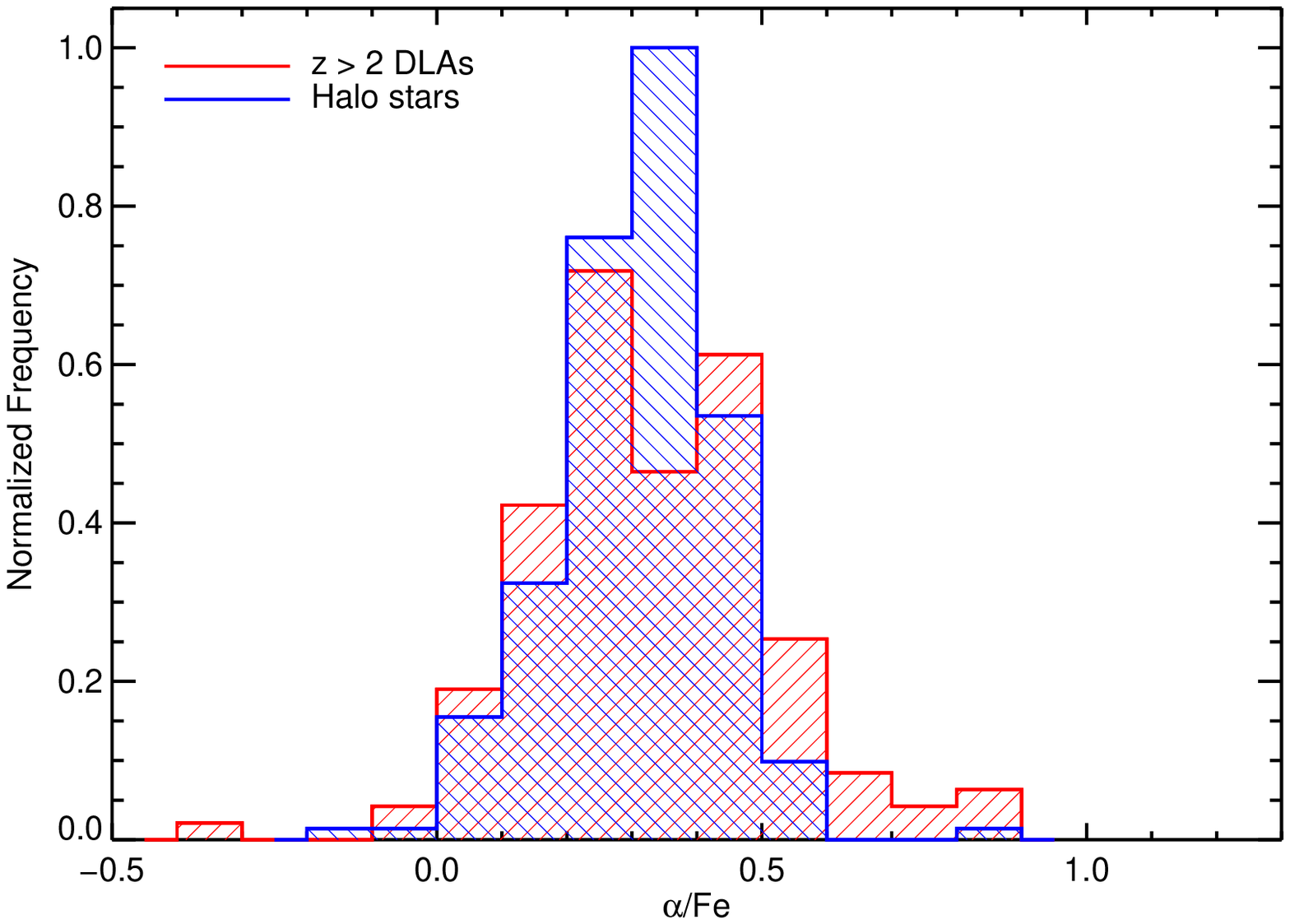}}
\hfill
\subfigure[$z>2$ and \mh ~$< -1$ DLAs]{\includegraphics[scale=0.48, viewport=15 5 495 360,clip]{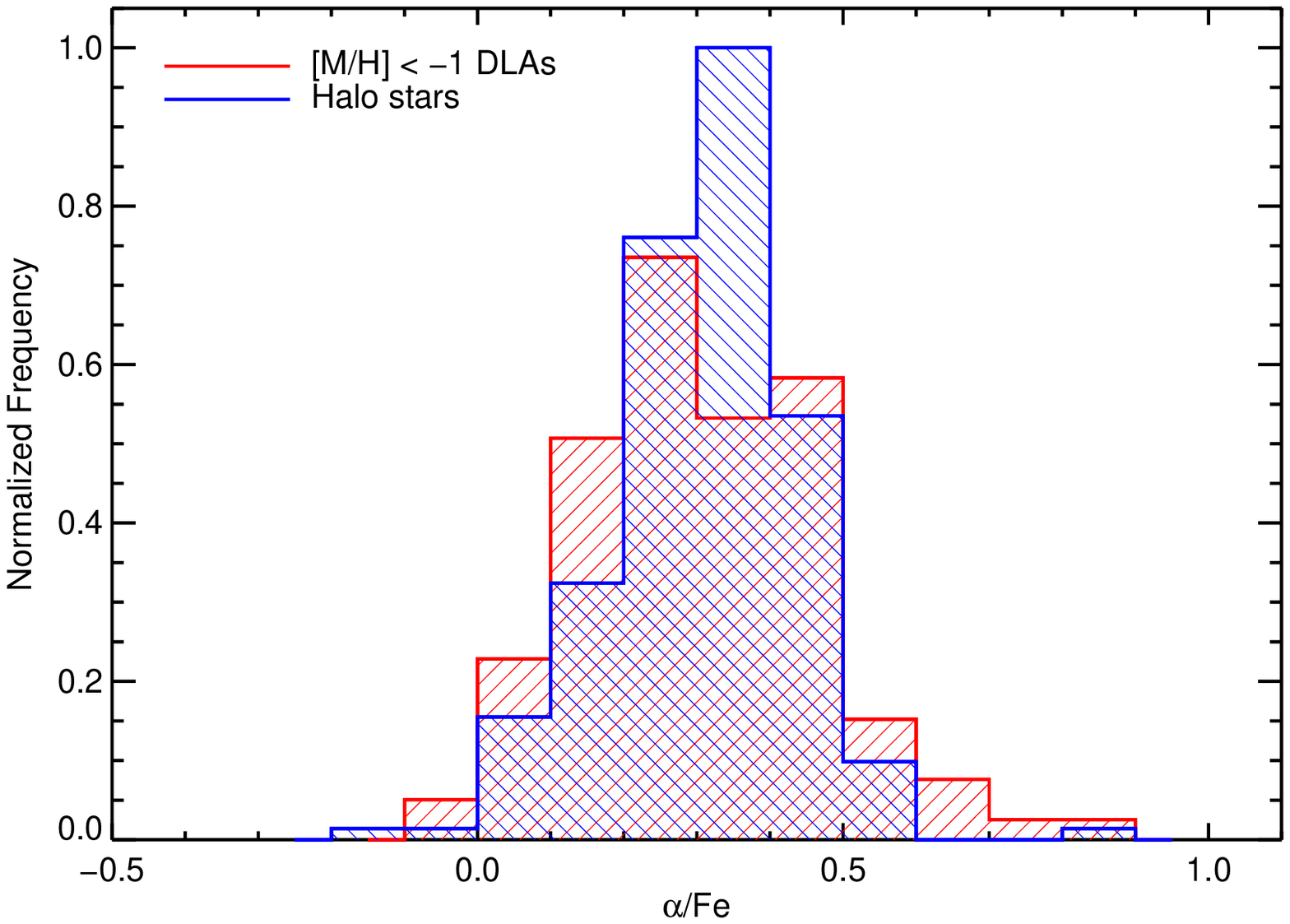}}
}
\caption{Histogram comparing $\alpha$/Fe ratios of 207 halo stars with a) 138 DLAs at $z>2$ and
b) 115 DLAs at $z>2$ with \mh ~$<-1$.
The DLAs are selected such that they have both $\alpha$ and Fe elemental abundances measured. 
The histograms are normalized to have equal area and a maximum value of 1.0.
The red left shaded region represents the DLAs and the blue right
shaded region represents the halo stars.  
The $\alpha$/Fe ratios of DLAs is consistent with that of halo stars. 
}
\label{fig:halofe}
\end{figure}

In Figure \ref{fig:halofe}, we plot histograms of the [$\alpha$/Fe] ratios for the 207 halo stars (blue) and 
a) 138 $z>2$ DLAs that have abundances for both an Fe and an $\alpha$-element, and b) a subset 115 of those 138
DLAs that also have [M/H] $<-1$. 
We note that it is possible that the Fe abundances are depleted onto dust grains, especially for the higher metallicity 
systems. However, at lower metallicities, the [$\alpha$/Fe] ratio of DLAs is relatively constant \citep[see Figure \ref{fig:alphafe} and][]{Prochaska:2002b, Wolfe:2005}.
If there was dust depletion in the sample, it would move the red histogram to the right in Figure \ref{fig:halofe}a. 
The sample in Figure \ref{fig:halofe}b reduces such depletion issues, as the lower metallicity DLAs will have Fe less or not at all depleted (see Figure \ref{fig:alphafe}). 
In comparison, the mean alpha enhancements of the thin disk and thick disk are $\sim0.16$ and $\sim0.20$ respectively,  inconsistent with those of DLAs. 

We apply the K-S test to the [$\alpha$/Fe] ratios for DLAs and halo stars, and find the probability that the two distributions are drawn 
from the same parent population to be 3\% for the 138 DLAs in Figure \ref{fig:halofe}a, 
and  40\% for the 115 DLAs in Figure \ref{fig:halofe}b. We therefore 
cannot reject the null hypothesis with more than 1$\sigma$ confidence,
and find that the [$\alpha$/Fe] ratio distribution for the DLA sample and the halo stars
are compatible. This suggests that both the halo stars and the DLA gas are $\alpha$-enhanced 
by $\approx0.3$ dex, suggesting that halo stars could be formed out of DLA gas. 
This result is different than some previous studies \citep[e.g.][]{Pettini:1999, Centurion:2000, Molaro:2000, Ledoux:2002a, Vladilo:2002, Nissen:2004, Nissen:2007} mainly due to a 
different interpretation of the behavior of Zn (see \S\ref{order} and \S\ref{comp}), and previous small number statistics. 
For instance, \citet{Nissen:2004} find no [S/Zn] enhancement in DLAs, and therefore conclude that there is no 
$\alpha$-enhancement. However, we find that Zn behaves like an $\alpha$-element, 
and therefore does not constrain the [$\alpha$/Fe] ratio.

\vspace{-5mm}

\section{Discussion}
\label{discus}
\subsection{Metallicity Evolution}
\label{discuss_metalevol}
The principal result of this study is that we find a continued decrease in the metallicity of DLAs with increasing
redshift out to $z\approx5$. We improve the significance of this trend to $6.7\sigma$ and extend it to higher redshifts.
These observations of the metallicity evolution tightly constrain the star formation history of $z \sim 5$ galaxies
and the processes that transport metals from star-forming regions to the ambient ISM. 

\begin{figure}[b!]
\center{
\includegraphics[scale=0.45, viewport=15 5 495 360,clip]{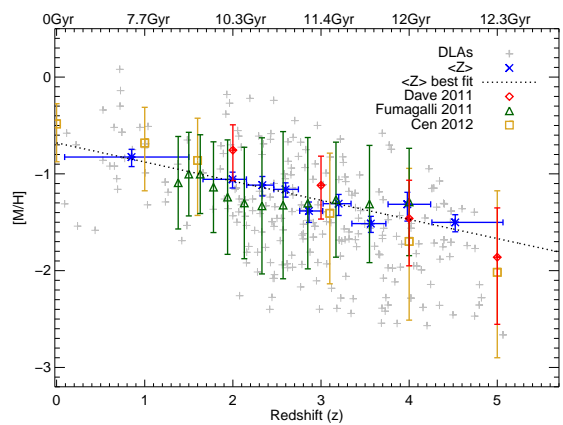}
}
\caption{Comparison of the DLA metal abundance evolution from simulations to the data. 
The gray plus signs are all DLAs used in this study, the blue crosses with error bars show the cosmic metallicity,
{\Z}, and the black straight line is a linear fit to the {\Z} data points. The red diamonds are mean metallicities of DLAs extracted
from the simulation presented in \citet{Dave:2011}. The green triangles by models from \citet{Fumagalli:2011} based on simulations
by  \citet{Ceverino:2010}. The gold squares are from \citet{Cen:2012}. All the error bars are 1$\sigma$ confidence levels. The
normalization of the metal abundances all agree with the data, while the slope in the evolution of the metal abundance is somewhat different,
although consistent within the uncertainties. 
}
\label{fig:metal_model}
\end{figure}

Figure \ref{fig:metal_model} compares current models to our measured abundance evolution. 
We show the same data and best fit cosmic metallicity as in Figure \ref{fig:metal}, and over-plot three models from simulations. 
The red diamonds are mean metallicities of DLAs extracted from the 48 Mpc/h momentum-driven wind simulation presented in \citet{Dave:2011}, 
using the self-shielding criterion in \citet{Popping:2009}, including an molecular fraction calculation based on \citet{Blitz:2006}. The green triangles are from
\citet{Fumagalli:2011} based on simulations by \citet{Ceverino:2010} using an atomic fraction calculation based on the formalism by \citet{Krumholz:2009jb}.
Lastly, the gold squares are directly from \citet{Cen:2012}. 

The normalization of the metal abundance of DLA gas in the simulations are remarkably similar to the observed values for all three models. In addition, the evolution of the 
metal abundance is very similar for two of them. The models by \citet{Dave:2011} and \citet{Cen:2012} agree within their uncertainties, although they strictly do have
steeper slopes than the data. On the other hand, the models by \citet{Fumagalli:2011} show no real evolution in the DLA metallicity, which may be due to sampling a small number of halos, or other issues in the simulation. In all, it is reassuring that different models with different hydro-formulations seem to broadly agree on the metal content, and match the measured values. 
However, although these models treat metal diffusion, mixing, and feedback differently, the broad agreement between the models suggests that perhaps the metals are mostly locked in the high density gas where stars are produced (private communication, Michele Fumagalli 2012). 

In addition to the continued decrease in metallicity, 
Figure \ref{fig:metal} reveals two interesting features. The first is the large 
dispersion of  [M/H] at $z>4$ that matches the results at lower redshift \citep{Prochaska:2003b}. 
This result is not an observational uncertainty but is due to intrinsic scatter amongst the DLAs.
The scatter does not appear to evolve with redshift, suggesting that it is intrinsic to the 
DLAs, possibly due to the wide range of masses of the galaxies hosting DLAs (Neeleman et al. 2012 in prep),
as suggested by the mass-metallicity correlation found both at low redshift \citep{Tremonti:2004}
and at high redshift \citep{Erb:2006,Prochaska:2008}.

The second feature is a metallicity ``floor'' at [M/H]$\approx$$-$3, which is also consistent 
with the results at lower redshift \citep{Prochaska:2003b,Penprase:2010}. 
We note that we cannot currently rule out the alternative possibility that rather than a floor,
this cutoff is due to the tail of the presumably gaussian distribution not being adequately sampled
(see Figure \ref{fig:metalhist}). This is investigated in 
\citet{Penprase:2010}, who find that if it is the gaussian tail, the probability of finding a DLA with [M/H]$\lesssim-3$ is 
extremely small. 

If the observed metallicity ``floor'' is real, then it is likely a physical lower limit of the metallicity of DLAs out to at least $z\approx5$, which may continue
out to even larger redshifts.  
We note that if there is gas with very low metallicity, it has a very small cross-section at {\nh}$\geq 2\times10^{20}$cm$^{-2}$.
The observed DLA metallicities are systematically higher than those of the {\lya} forest 
clouds which have [M/H]=$-$3 \citep{Aguirre:2004}, and therefore even the lowest metallicity
DLA gas is part of a distinctly different population. Moreover, 
the extension of this lower limit of [M/H] at higher redshift has implications for the primordial gas
in these galaxies. Either the primordial gas does not exist in the neutral phase in high redshift galaxies,
or always coincides with a metal enriched region \citep{Prochaska:2003b}. Simulations of primordial metal enrichment 
including Population III stars show that a single pair-instability supernova can enrich a halo to [M/H]$\approx$$-$3 \citep{Wise:2012},
which would explain and be consistent with the observed ``floor''. Such a pre-enrichment of the ISM by Popullation III stars resulting 
in a minimum metallicity would also improve the agreement of simulated galaxies and low-mass dwarf galaxies \citep{Tassis:2012}. 

\begin{figure*}
\center{
\includegraphics[scale=0.75, viewport=15 5 495 360,clip]{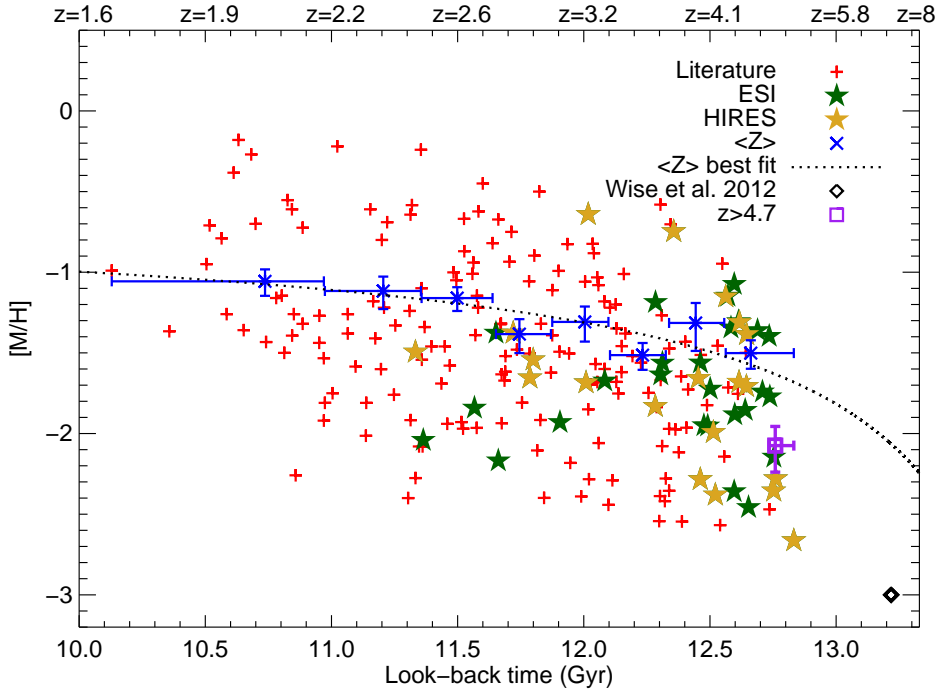}
}
\caption{DLA metal abundance [M/H] versus time similar to figure \ref{fig:metal}. The black diamond is the prediction from a 
hydrodynamic simulation on primordial metal enrichment \citep{Wise:2012}. The purple square is a column density weighted 
mean of the 7 highest redshift DLAs at $z>4.7$ in our sample, which all have values below the fit. 
Red plus signs from the literature. 
The metallicities from ESI are green stars and those from HIRES are gold stars.
The 9 blue data points with error bars show cosmic metallicity,
{\Z}, where horizontal error bars are determined such that there are equal numbers
of data points per redshift bin. Black straight line is linear fit to
{\Z} data points as a function of redshift.
We note that we show only the values at $z>1.6$ to focus on the area of interest at higher redshift. 
}
\label{fig:metal_time}
\end{figure*}

While the evolution of {\Z} is linear in redshift, it is non-linear
in time, with the slope, d{\Z}/dt, becoming significantly steeper at earlier cosmic times.
In Figure \ref{fig:metal_time}, we plot the metallicity as a function of time rather than redshift. We truncate the plot 
at $z=1.6$ in order to focus on the higher redshift part of the plot which contains the most rapidly changing metal abundances.
The black dotted line is the linear fit to {\Z} versus redshift, shown as a function of time,
and clearly shows the steepening of {\Z} going back in time.  While this trend line is not a physical model, 
it shows the importance of obtaining metallicities of the highest redshift DLA systems, as undertaken in this study.
Moreover, we show the metallicity prediction from simulations of primordial metal enrichment as a black diamond \citep{Wise:2012},
which would require a drop in metallicity below what is extrapolated from the data.

An intriguing property of the [M/H] values in Figure \ref{fig:metal} is that the 
highest redshift DLAs in our sample appear to have metallicities below the 
linear trend in {\Z}. Specifically,  6 of the 7 DLAs at $z>4.7$  have [M/H] values below the linear fit ({\Z}$=-1.6$ at $z=4.8$), with 
a column density weighted mean {\Z} $= -2.1^{+0.1}_{-0.2}$ (The DLA that does not is the lowest redshift of the group). 
We note that the linear fit is not a physical model, and therefore does not constrain our expectations.
However, the physical models shown in Figure \ref{fig:metal_model} do not currently show a sudden drop in metallicity, so if real, this would mark 
a departure of these models from the data. 
Although the sample size is currently small, and this result may just be 
due to small number statistics, future work should test whether this departure of {\Z} 
below the linear fit persists at at $z>4.7$. 

\subsection{Comparison with Halo Stars}
\label{halocomp}

The metallicity measurements at high redshift also enable us to investigate
whether any known stellar populations in our Galaxy have the same metallicity distribution
as the high redshift DLAs. Previously, no known population
was known to have the same metallicity distribution as DLAs \citep{Pettini:1997, Pettini:2004, Pettini:2006}. 
This was because the metallicity of thin and thick disks were too metal rich for DLAs,
while the halo stars were too metal poor. 
However, we showed in \S\ref{halo} that it is possible that the metallicity distribution of 
DLAs at $z>2$ is compatible with that of halo stars. 
Specifically, we cannot rule out, with a high degree of confidence, that DLAs are the progenitors of halo stars. 
We stress that this is the first time 
that the metallicity distribution of any known population of stars have been shown 
to be consistent with being drawn from the same parent population as DLAs. 

There are two primary differences between our study and previous studies, which compared the DLA metallicity 
distributions to known stellar populations \citep{Pettini:1997, Pettini:2004, Pettini:2006}.
First, we use $\alpha$-elements for abundance measurements (primarily S and Si), while
the other studies use Zn. Second, we probe a higher redshift range out to $z\sim5$, as opposed to $z\sim3.5$. 
Both Zn and Si yield reliable abundance measurements, which are consistent with each other (see \S\ref{comp}),
although, the oscillator strengths of Zn are smaller than Si, making it much more difficult to detect low metallicity DLAs.
Additionally, the wavelengths of Zn transitions are larger than Si, making Zn measurements at high
redshift more difficult to obtain as they are redshifted into the near-infrared. 
The past comparisons of the metallicity distributions included upper limits of Zn together with detections, skewing the
distribution towards metallicities higher than that of the intrinsic population. Furthermore, by sampling higher redshifts, we include
lower metallicity DLAs, since the metallicity of DLAs evolve with time. Together, our sensitivity to lower metallicity systems 
and our inclusion of higher redshift systems account for the different metallicity distribution presented here.

In addition to the metallicity distribution, the chemistry of the DLAs and halo stars is consistent,
based on the comparison of the [$\alpha$/Fe] ratios. Both populations have enhanced
$\alpha$-elements ([$\alpha$/Fe] $\approx$ 0.3), suggesting that stars formed shortly after the ISM was enriched by 
type~II supernovae, but before the type Ia supernovae had a chance to go off \citep{Matteucci:2003, Venn:2004}.
Together, these results suggest that the halo stars formed out of gas that has undergone the same number of generations of star formation 
as DLAs, which occurred before type Ia supernovae contributed significantly. It is therefore possible that the halo stars in
the Milky Way formed out of $z>2$ DLA gas.

There are two primary formation scenarios for the Galactic halo stars  (for a more thorough review, see \citet{Helmi:2008}). 
The first is an in-situ scenario with the formation of halo stars from in-falling gas towards the Galactic center, which later collapse onto the plane 
\citep[e.g.][]{Eggen:1962, Samland:2003}. 
The second is an accretion scenario where a number of smaller galaxies merge to form the Galactic halo 
\citep[e.g.][]{Searle:1978, Bullock:2005, Moore:2006}. 
The later is often considered the more favored scenario, due to its resemblance to the model of hierarchical assembly with the collapse of small haloes of cold dark matter (CDM) \citep{White:1978}. 
It is likely that the answer lies somewhere in-between, with both processes playing a role \citep{Majewski:1993, Chiba:2000, Helmi:2008, Font:2011ca}.
In addition, while Galactic halo stars used to be considered a single stellar population, recent evidence suggests 
the existence of an inner and an outer halo \citep{Carollo:2007, Lai:2009, Carollo:2010}. The formation mechanism of these two populations are likely different;
with the inner halo  formed from a combination of accretion of early rapid mergers and in-situ formation, and the outer halo formed purely from the disruption of lower mass satellite galaxies and accretion \citep{Carollo:2007, Zolotov:2009, Carollo:2010, Cooper:2010}.

The Galactic stellar halo formation scenarios are compatible with halo stars formed out of DLA gas. Specifically, these scenarios include the mergers of 
lower mass systems that form the majority of the stars. While nearby dwarf spheroidal 
galaxies were originally thought to lack stars with [Fe/H] $\sim-3$ \citep{Helmi:2006},
fainter systems have  sufficiently low metallicity that they could be 
contributors to the Galactic halo  \citep{Kirby:2008fs, Koch:2008hw,Norris:2008, Frebel:2010cg, Lai:2011}.
Such low metallicity stars are even now seen in more massive dwarf spheroidal galaxies \citep{Kirby:2011bm}.  
In addition, while initially the $\alpha$-enhancement of the dwarf galaxies did not match the stellar halo \citep{Shetrone:2003de,Geisler:2007kw}, 
there is now agreement for low metallicity stars in such galaxies \citep{Frebel:2010cg, Kirby:2011ca}.
Similar to dwarf galaxies that may have been halo progenitors, the $\alpha$-enhancement and metallicity distribution of DLAs at $z>2$ match the halo stars (see Figures \ref{fig:halo} and \ref{fig:halofe}).
In fact, DLAs encompass both lower mass galaxies 
and the outer regions of spiral galaxies \citep{Matteucci:1997, Prochaska:1998, DessaugesZavadsky:2004, DessaugesZavadsky:2007, Meiring:2011iz, Fumagalli:2011}
and therefore a subset of DLAs are in fact dwarf galaxies (although not all DLAs can be dwarf galaxies \citep{Prochaska:2005}). 
It is therefore not overly surprising that DLAs also share characteristics with halo stars, and that halo stars may have formed out of DLA gas.\footnote{
We note that if the halo stars formed out of DLA gas, then the
higher kinematics of halo stars than DLAs 
would necessarily be caused by mergers, which randomize and increase the kinetic energy
of the orbits of the stars formed in the quiescent DLAs.}

\subsection{Star Formation Rate Efficiency}

The metallicity evolution also has implications for the star formation rate (SFR) efficiency at high redshift. 
In general, the SFR per unit area is related to the gas density via the locally established 
Kennicutt-Schmidt (KS) relation ($\Sigma_{SFR}$ $\propto$ $\Sigma_{gas}^{\beta}$). 
At $z\sim3$, the {\it in situ} SFR of DLA gas was found to be less
than 5$\%$ of what is expected from the KS relation in isolated
regions \citep{Wolfe:2006}, and $\sim10\%$ of the expected rate in
the outskirts of LBGs \citep{Rafelski:2009, Rafelski:2011}.
This means that the SFR of DLA gas is less efficient than what would be
predicted by the KS relation. \citet{Gnedin:2010} 
conclude that the main reason for the decreased efficiency of star formation is that the diffuse ISM in high
redshift galaxies has a lower metallicity and dust-to-gas ratio, both
of which are needed to cool the gas and form stars.  
In this case, since the metallicities of DLAs decrease with redshift,
we expect that the efficiency of star formation should be correlated
with redshift.  In order to further our understanding of the effect of
metallicity on the SFR efficiency, and measure the evolution of the KS
relation with redshift, measurements over a range in redshift are required. 
This study provides the metallicities of the DLA gas over a large redshift range, 
setting the stage for such studies.

\section{Summary}
\label{sum}

The purpose of this study is to measure the metallicities of DLAs with $z>4$ in order to 
determine the metallicity evolution of DLAs out to $z\sim5$.
We obtained high resolution spectroscopy of 68 quasars 
using the ESI and HIRES instruments on the Keck Telescopes
and find the following: \\

\noindent (1) We fit {\nh} Voigt profiles to 68 candidate DLAs from SDSS, and confirm 51 of them to have
{\nh}$\geq 2\times10^{20}$cm$^{-2}$ (see Figures \ref{fig:rogue0} and \ref{fig:rogue1}). 
For the 47 $z>4$ DLA candidates, 15 are false positives leaving 32 confirmed DLAs with $z>4$ (see Figure \ref{fig:multi}), tripling 
the sample of $z>4$ DLAs that have high resolution metallicity measurements to a total of 39. 
While at lower redshifts the majority of candidate
DLAs from SDSS are confirmed, at $z>4$ the number of false positives is large, with $\sim$30\% misidentified due to 
line-blending with random {\lya} forest clouds.  Studies using DLA identifications from SDSS at $z>4$ therefore need to 
apply a correction for the increased false positive rate for the lowest {\nh} DLAs
before determining the column density distribution function, the total 
covering fraction, and the integrated mass density of \ion{H}{1} gas. \\

\noindent (2) We consider the different elements available to measure the metallicity of DLAs, and find that the metallicity is determined
with the highest confidence from elements in the following order: O, S, Si, Zn, Fe. 
We compare the elemental abundances determined from different ions in Figure \ref{fig:comp}, and conclude that S, Si, and Zn
all result in similar metallicities. In addition, we find that for low metallicity DLAs, the metallicity can be determined
from Fe, with a correction due to the alpha enhancement of DLAs. 
In Figures \ref{fig:alphahist} and \ref{fig:alphafe}, we show that DLAs are alpha enhanced, with a mean value of $0.27\pm0.02$. \\

\noindent (3) We measure metal abundances, [M/H], for a total of 47 DLAs (see Table \ref{metal:tab2} 
and Figure \ref{fig:metalobs}), with 30 at $z>4$ (one of which is a proximate DLA). 
We find that the cosmic metallicity,  {\Z},
continues to decrease with increasing redshift to $z\approx5$ (see Figure \ref {fig:metal}). 
Specifically, we find that $\langle {\rm Z}\rangle=(-0.22 {\pm} 0.03)z-(0.65 {\pm}0.09)$, which is a $6.7\sigma$ detection in 
the evolution of {\Z} and is a significant improvement
over the previous $\sim3\sigma$ detections by \citet{Prochaska:2003b}, \citet{Kulkarni:2005}, \citet{Kulkarni:2007}, and \citet{Kulkarni:2010}. 
If we remove our lowest redshift bin which may be slightly metallicity biased, and thereby only using DLAs with $z>1.5$, we find that  
$\langle {\rm Z}\rangle=(-0.22 {\pm} 0.05)z-(0.66 {\pm}0.15)$, which is a $4.6\sigma$ detection in the evolution of {\Z}.
The fit values are completely consistent with the previous measurements by \citet{Prochaska:2003b},
confirming that the mean metallicity of the universe in neutral gas is doubling about every billion years at $z\sim3$. 
Additionally, the normalization and evolution of the metal abundance of simulations agree remarkably well with our data, 
as shown in Figure \ref{fig:metal_model}. \\

\noindent (4) We find that the large dispersion of  [M/H] of $\sim0.5$ dex measured at $z\lesssim4$ \citep{Prochaska:2003b}
continues out to $z\sim5$ (see Figure \ref{fig:metalobs}). This result is not an observational uncertainty, 
but is due to intrinsic scatter amongst the DLAs.
The scatter does not appear to evolve with redshift, and could be due to the wide range of masses of the galaxies hosting DLAs
(Neeleman et al. 2012 in prep). \\

\noindent (5) We find that the metallicity ``floor'' at [M/H]$\approx$$-$3 measured at $z\lesssim4$ \citep{Prochaska:2003b}
continues out to $z\sim5$ (see Figure \ref{fig:metalobs}). 
Stated differently, we find no DLAs with [M/H] $<$ $-$2.8 despite the sensitivity of our spectra for finding DLAs with [M/H] $\ge -4.5$.
This is likely a physical lower limit of the metallicity of DLAs out to at least $z\approx5$, which may continue out to even larger redshifts.  \\

\noindent (6) We find that the metallicity distribution and the $\alpha$/Fe ratios of $z>2$ DLAs is consistent with being drawn
from the same parent population as halo stars 
(see Figures \ref{fig:halo} and \ref{fig:halofe}). This is the first time that the metallicity distribution of any known stellar population
has been shown to be consistent with being drawn from the same parent population as DLAs. 
It is therefore possible that the halo stars in the Milky Way formed out of gas that commonly exhibits DLA absorption at $z>2$. \\

Altogether, we have measured the evolution of the metal abundances of DLAs out to $z\sim5$, 
which tightly constrain the star formation history of $z \sim 5$ galaxies
and the processes that transport metals from star-forming regions to the ambient ISM. 
The 6 highest redshift DLAs in our sample all have both [M/H] and {\Z} values below the linear fit. 
Future observations should therefore focus on obtaining metallicities at $z>4.7$ to further constrain the metallicity evolution of DLAs, 
as they will have the largest impact. 
Lastly, we find that the metallicity distribution and $\alpha$-enhancement of DLAs is similar to Galactic halo stars, enabling the possibility that 
halo stars form out of DLA gas. 

\acknowledgements

The authors thank David Lai and Max Pettini for helpful conversations on halo-stars.
Support for this work was provided by NSF grant AST 07-09235 and the Chancellor's associates fund at UCSD.
The W. M. Keck Observatory is operated as a scientific partnership among the California Institute 
of Technology, the University of California and the National Aeronautics and 
Space Administration.  The Observatory was made possible by the generous 
financial support of the W. M. Keck Foundation.  The authors wish to recognize and 
acknowledge the very significant cultural role and reverence that the summit 
of Mauna Kea has always had within the indigenous Hawaiian community.  We are 
most fortunate to have the opportunity to conduct observations from this 
mountain.  

{\it Facility:} 
 \facility{Keck:II (ESI)} \facility{Keck:I (HIRES)}

\begin{appendices}
\input{tab4.tex}

\vspace{-20mm}
\section{ONLINE MATERIAL}

Table \ref{metal:tab4} provides the column density measurements for the entire new DLA sample.
Column (1) gives the quasar coordinate name obtained from the SDSS survey, 
column (2) gives the DLA absorption redshifts determined from the metal transition lines. 
Column (3) gives the Ion measured, column (4) gives the wavelength of said ion, 
column (5) gives the oscillator strength used, and column (6) gives the instrument used.
Column (7) gives the velocity interval over which the equivalent width and column density are measured,
and column (8) gives the rest frame equivalent width. 
Column (9) gives the measured column density, and column (10) gives the adopted column density value
combining measurements fo multiple transitions.

Figure set 17 shows the velocity profiles of element transitions for the DLAs in Table \ref{metal:tab4}. Every column density measurement
in this table is shown in these Figures. These velocity profiles are a mix of HIRES and ESI data, as specified in Table \ref{metal:tab4}, 
although it should be quite obvious based on the resolution of the data in the Figures.

\input{figset17.tex}

\end{appendices}

\pagebreak
\clearpage

\bibliography{refs} 

\end{document}

%% file: tab1.tex
\begin{deluxetable*}{rrrccrrcccrccccccccccccc}
\tabletypesize{\scriptsize}
\tablecaption{Journal of Observations
\label{metal:tab1}}
\tablewidth{0pt}
\tablehead{
\colhead{QSO} &
\colhead{R.A.} &  
\colhead{Dec.} & 
\colhead{$r$} &
\colhead{$i$} &
\colhead{$z_{em}$} &
\colhead{$z_{abs}$} &
\colhead{Date} & 
\colhead{Instr.} & 
\colhead{Slit} & 
\colhead{$t_{exp}$} &
\colhead{S/N\tablenotemark{a}} \\
\colhead{} &
\colhead{(J2000)} &  
\colhead{(J2000)} & 
\colhead{(AB)} &
\colhead{(AB)} &
\colhead{} &
\colhead{} &
\colhead{(UT)} & 
\colhead{} & 
\colhead{(arcsec)} & 
\colhead{(s)} &
\colhead{pixel$^{-1}$} \\
\colhead{(1)} &
\colhead{(2)} &  
\colhead{(3)} & 
\colhead{(4)} &
\colhead{(5)} &
\colhead{(6)} &
\colhead{(7)} &
\colhead{(8)} & 
\colhead{(9)} & 
\colhead{(10)} & 
\colhead{(11)} &
\colhead{(12)}
}

\startdata

J0040$-$0915 & 00 40 54.7 & $-$09 15 27 & 20.48 & 19.18 & 4.98 & 4.74 & 2010 Jan 20 & ESI & 0.50 & 2280 & 16 \\
& & & & & & & 2011 Jan 16 & HIRESr & 0.86 & 7200 & 6 \\
& & & & & & & 2011 Jan 24 & HIRESr & 0.86 & 3600 &  \\
J0210$-$0018 & 02 10 43.2 & $-$00 18 18 & 20.48 & 19.20 & 4.72 & 4.57\tablenotemark{b} & 2010 Jan 20 & ESI & 0.50 & 2880 & 9 \\
J0331$-$0741 & 03 31 19.7 & $-$07 41 43 & 20.56 & 19.12 & 4.71 & 4.19\tablenotemark{b} & 2010 Jan 20 & ESI & 0.75 & 4320 & 18 \\
J0747+4434 & 07 47 49.7 & +44 34 17 & 19.62 & 17.35 & 4.43 & 4.02 & 2009 Mar 22 & ESI & 0.75 & 3600 & 15 \\
& & & & & & & 2011 Jan 16 & HIRESr & 0.86 & 3601 & 9 \\
& & & & & & & 2011 Jan 24 & HIRESr & 0.86 & 10800 &  \\
J0759+1800 & 07 59 07.6 & +18 00 55 & 21.06 & 19.16 & 4.86 & 4.66 & 2010 Jan 20 & ESI & 0.75 & 4320 & 18 \\
J0813+3508 & 08 13 33.3 & +35 08 11 & 20.81 & 19.13 & 4.92 & 4.30\tablenotemark{b}  & 2010 Apr 21 & ESI & 0.75 & 1400 & 14
\enddata

\tablecomments{Units of right ascension are in hours, minutes,
  and seconds, and units of declination are in degrees,
  arcminutes, and arcseconds. Photometry from SDSS \citep{Abazajian:2009}.This table is available in its entirety in a machine-readable form in the online journal. A portion is shown here for guidance regarding its form and content.}
\tablenotetext{a}{Median signal-to-noise per pixel at $\lambda\approx 8300$\AA, except in the HIRES observations of J1353+5328 and J1541+3153, where it is at $\lambda\approx 7300$ \AA.}
\tablenotetext{b}{Marked as a DLA in SDSS, but is not a DLA
  based on higher resolution ESI data}

\end{deluxetable*}

%% file: tab2.tex
\begin{deluxetable*}{rrrcrcrcrccccccccc}
\tabletypesize{\scriptsize}
\tablecaption{New DLA metallicities
\label{metal:tab2}}
\tablewidth{0pt}
\tablehead{
\colhead{QSO} &
\colhead{$z_{abs}$} &  
\colhead{log$N_{\rm HI}$} & 
\colhead{$f_{\rm \alpha}$\tablenotemark{a}} &
\colhead{[$\alpha$/H]} &
\colhead{$f_{\rm Fe}$\tablenotemark{b}} &
\colhead{[Fe/H]} & 
\colhead{$f_{\rm mtl}$\tablenotemark{c}} &
\colhead{[M/H]} \\
\colhead{(1)} &
\colhead{(2)} &  
\colhead{(3)} & 
\colhead{(4)} &
\colhead{(5)} &
\colhead{(6)} &
\colhead{(7)} &
\colhead{(8)} & 
\colhead{(9)} 
}

\startdata
J0040$-$0915 &4.7394 &20.30$ \pm $0.15 &0 &\nodata &1 &$-1.70\pm $0.16 &2 &$-1.40\pm $0.17 \\
J0747+4434 &4.0196 &20.95$ \pm $0.15 &3 &$<-1.90\pm $0.15 &4 &$-2.58\pm $0.20 &2 &$-2.28\pm $0.20 \\
J0759+1800 &4.6577 &20.85$ \pm $0.15 &4 &$-1.74\pm $0.16 &3 &$<-1.14\pm $0.15 &1 &$-1.74\pm $0.16 \\
J0817+1351 &4.2584 &21.30$ \pm $0.15 &4 &$-1.15\pm $0.15 &1 &$-1.30\pm $0.16 &1 &$-1.15\pm $0.15 \\
J0825+3544 &3.2073 &20.30$ \pm $0.10 &2 &$>-1.83\pm $0.10 &1 &$-1.98\pm $0.10 &2 &$-1.68\pm $0.16 \\
J0825+3544 &3.6567 &21.25$ \pm $0.10 &1 &$-1.83\pm $0.13 &4 &$-1.95\pm $0.11 &1 &$-1.83\pm $0.13 \\
J0825+5127 &3.3180 &20.85$ \pm $0.10 &1 &$-1.67\pm $0.14 &1 &$-2.08\pm $0.10 &1 &$-1.67\pm $0.14 \\
J0831+4046 &4.3440 &20.75$ \pm $0.15 &1 &$-2.36\pm $0.15 &1 &$-2.41\pm $0.17 &1 &$-2.36\pm $0.15 \\
J0834+2140 &3.7102 &20.85$ \pm $0.10 &3 &$<-1.69\pm $0.10 &1 &$-1.86\pm $0.10 &2 &$-1.56\pm $0.16 \\
J0834+2140 &4.3900 &21.00$ \pm $0.20 &4 &$-1.30\pm $0.20 &1 &$-1.69\pm $0.20 &1 &$-1.30\pm $0.20
\enddata

\tablecomments{Note that none of the reported limits take into account the uncertainty in $N_{\rm H I}$. This table is available in its entirety in a machine-readable form in the online journal. A portion is shown here for guidance regarding its form and content.}
\tablenotetext{a}{0 = No measurement; 1= Si measurement; 2 = Si lower limit; 3 = Si upper limit; 4 = S measurement, .}
\tablenotetext{b}{0 = No measurement; 1= Fe measurement; 2 = Fe lower limit; 3 = Fe upper limit; 
4 = [Ni/H]$-$0.1 dex; 5 = [Cr/H]$-$0.2 dex; 6 = [Al/H].}
\tablenotetext{c}{1 = [$\alpha$/H]; 2=[Fe/H]. In the latter case, we use [M/H] = [Fe/H]+0.3 dex.}
\tablenotetext{d}{This is a proximate DLA, and is therefore not included in the analysis of the metallicity evolution.}

\end{deluxetable*}

%% file: tab3.tex
\begin{deluxetable*}{lrrcrcrcrrcccccccc}
\tabletypesize{\scriptsize}
\tablecaption{Literature DLA metallicities
\label{metal:tab3}}
\tablewidth{0pt}
\tablehead{
\colhead{QSO} &
\colhead{$z_{abs}$} &  
\colhead{log$N_{\rm HI}$} & 
\colhead{$f_{\rm \alpha}$\tablenotemark{a}} &
\colhead{[$\alpha$/H]} &
\colhead{$f_{\rm Fe}$\tablenotemark{b}} &
\colhead{[Fe/H]} & 
\colhead{$f_{\rm mtl}$\tablenotemark{c}} &
\colhead{[M/H]} &
\colhead{Refs\tablenotemark{d}}\\
\colhead{(1)} &
\colhead{(2)} &  
\colhead{(3)} & 
\colhead{(4)} &
\colhead{(5)} &
\colhead{(6)} &
\colhead{(7)} &
\colhead{(8)} & 
\colhead{(9)} & 
\colhead{(10)} 
}

\startdata

Q2359$-$02 &2.0951 &20.70$ \pm $0.10 &1 &$-0.72\pm $0.10 &1 &$-1.61\pm $0.10 &1 &$-0.72\pm $0.10 &10,22 \\
Q2359$-$02 &2.1539 &20.30$ \pm $0.10 &1 &$-1.53\pm $0.10 &1 &$-1.83\pm $0.10 &1 &$-1.53\pm $0.10 &10,22 \\
Q0000$-$2619 &3.3901 &21.41$ \pm $0.08 &5 &$-1.68\pm $0.20 &1 &$-2.11\pm $0.09 &1 &$-1.68\pm $0.20 &3,10,12,20 \\
PSS0007+2417 &3.7045 &20.55$ \pm $0.15 &2 &$>-1.69\pm $0.15 &4 &$-1.57\pm $0.26 &3 &$-1.27\pm $0.27 &34 \\
PSS0007+2417 &3.4960 &21.10$ \pm $0.10 &1 &$-1.52\pm $0.11 &4 &$-1.73\pm $0.11 &1 &$-1.52\pm $0.11 &34 \\
PSS0007+2417 &3.8382 &20.85$ \pm $0.15 &6 &$-2.12\pm $0.28 &1 &$-2.39\pm $0.15 &4 &$-2.12\pm $0.24 &34 \\
Q0010$-$002 &2.0250 &20.80$ \pm $0.10 &0 &\nodata &1 &$-1.28\pm $0.11 &2 &$-1.16\pm $0.12 &31 \\
J0013+1358 &3.2814 &21.55$ \pm $0.15 &1 &$-2.06\pm $0.16 &1 &$-2.66\pm $0.15 &1 &$-2.06\pm $0.16 &53 \\
Q0013$-$004 &1.9730 &20.83$ \pm $0.07 &4 &$-0.70\pm $0.08 &1 &$-1.47\pm $0.08 &1 &$-0.70\pm $0.08 &27,31 \\
BR0019$-$15 &3.4389 &20.92$ \pm $0.10 &1 &$-1.01\pm $0.11 &4 &$-1.53\pm $0.11 &1 &$-1.01\pm $0.11 &10,22 
\enddata

\tablecomments{This table is available in its entirety in a machine-readable form in the online journal. A portion is shown here for guidance regarding its form and content.}
\tablenotetext{a}{0 = No measurement; 1= Si measurement; 2 = Si lower limit; 3 = Si upper limit; 4 = S measurement, 5 = O measurement, 6 = combined alpha limit.}
\tablenotetext{b}{0 = No measurement; 1= Fe measurement; 2 = Fe lower limit; 3 = Fe upper limit, 4 = combined [Fe/H] limits; 
4 = [Ni/H]$-$0.1 dex; 5 = [Cr/H]$-$0.2 dex; 6 = [Al/H], 7 = combined Fe limit}
\tablenotetext{c}{1 = [$\alpha$/H]; 2=[Fe/H].  with [M/H] = [Fe/H]+0.3 dex; 3 = combined [$\alpha$/H] limits, 4 = combined [Fe/H] limits}
\tablenotetext{d}{References:  1:\citet{Wolfe:1994}, 2:\citet{Meyer:1995}, 3:\citet{Lu:1996}, 4:\citet{Prochaska:1996}, 5:\citet{Prochaska:1997}, 6:\citet{Boisse:1998}, 7:\citet{Lu:1998}, 8:\citet{Lopez:1999}, 9:\citet{Pettini:1999}, 10:\citet{Prochaska:1999}, 11:\citet{Churchill:2000}, 12:\citet{Molaro:2000}, 13:\citet{Petitjean:2000}, 14:\citet{Pettini:2000}, 15:\citet{Prochaska:2000}, 16:\citet{Rao:2000}, 17:\citet{Srianand:2000}, 18:\citet{DessaugesZavadsky:2001}, 19:\citet{Ellison:2001}, 20:\citet{Molaro:2001}, 21:\citet{Prochaska:2001a}, 22:\citet{Prochaska:2001b}, 23:\citet{Ledoux:2002a}, 24:\citet{Ledoux:2002b}, 25:\citet{Levshakov:2002}, 26:\citet{Lopez:2002}, 27:\citet{Petitjean:2002}, 28:\citet{Prochaska:2002a}, 29:\citet{Songaila:2002}, 30:\citet{Centurion:2003}, 31:\citet{Ledoux:2003}, 32:\citet{Lopez:2003}, 33:\citet{Prochaska:2003a}, 34:\citet{Prochaska:2003b}, 35:\citet{DessaugesZavadsky:2004}, 36:\citet{Khare:2004}, 37:\citet{Turnshek:2004}, 38:\citet{Kulkarni:2005}, 39:\citet{Akerman:2005}, 40:\citet{Rao:2005}, 41:\citet{DessaugesZavadsky:2006}, 42:\citet{Ledoux:2006}, 43:\citet{Meiring:2006}, 44:\citet{Peroux:2006}, 45:\citet{Rao:2006}, 46:\citet{DessaugesZavadsky:2007}, 47:\citet{Ellison:2007}, 48:\citet{Meiring:2007}, 49:\citet{Prochaska:2007}, 50:\citet{Nestor:2008}, 51:\citet{Noterdaeme:2008}, 52:\citet{Peroux:2008}, 53:\citet{Wolfe:2008}, 54:\citet{Jorgenson:2010}, 55:\citet{Meiring:2011}, 56:\citet{Vladilo:2011}
}

\end{deluxetable*}

%% file: tab4.tex
\begin{deluxetable*}{lllccccccc}[h!]
\tabletypesize{\scriptsize}
\tablewidth{0pc}
\tablecaption{Ionic Column Densities \label{metal:tab4}}
\tablehead{\colhead{QSO} & \colhead{$z_{\rm abs}$} & \colhead{Ion} & \colhead{$\lambda$} & \colhead{${\rm log} ~f$}
& \colhead{Instr.} 
& \colhead{$v_{int}^a$} 
& \colhead{$W_\lambda^b$} 
& \colhead{${\rm log} ~N$} & \colhead{${\rm log} ~N_{\rm adopt}$} \\
& & & (\AA) & & & (\kms) & (m\AA) & & \\
(1)& (2)& (3)& (4)&(5)&(6)&(7)&(8)&(9)&(10)}
\startdata
J0040-0915 & 4.7394\\
& &\ion{C}{2} &1334.5323 &$ -0.8935$&ESI&$[ -80, 100]$&$ 449.4 \pm   8.0$&$> 14.64$&$> 14.64$\\
& &\ion{C}{4} &1548.1950 &$ -0.7194$&HIRES&$[ -40,  70]$&$ 126.9 \pm  21.9$&$> 13.60$&$ 13.77 \pm 0.03$\\
& & &1550.7700 &$ -1.0213$&HIRES&$[ -40,  70]$&$  99.7 \pm   7.5$&$ 13.77 \pm 0.03$& \\
& &\ion{O}{1} &1302.1685 &$ -1.3110$&HIRES&$[ -40,  70]$&$ 258.5 \pm   7.5$&$> 14.91$&$> 14.91$\\
& &\ion{Al}{2} &1670.7874 &$  0.2742$&HIRES&$[ -40,  70]$&$ 189.7 \pm  18.1$&$> 12.97$&$> 12.97$\\
& &\ion{Si}{2} &1304.3702 &$ -1.0269$&HIRES&$[ -40,  70]$&$ 129.4 \pm   7.8$&$> 14.16$&$> 14.18$\\
& & &1526.7066 &$ -0.8962$&HIRES&$[ -40,  70]$&$ 229.5 \pm  22.1$&$> 14.18$& \\
& &\ion{Si}{4} &1393.7550 &$ -0.2774$&HIRES&$[ -40,  70]$&$ 182.7 \pm   9.8$&$ 13.49 \pm 0.05$&$ 13.34 \pm 0.04$\\
& & &1402.7700 &$ -0.5817$&HIRES&$[ -40,  70]$&$  51.4 \pm  10.4$&$ 13.19 \pm 0.08$& \\
& &\ion{Fe}{2} &1608.4511 &$ -1.2366$&HIRES&$[ -30,  70]$&$ 103.0 \pm  13.1$&$ 14.05 \pm 0.06$&$ 14.05 \pm 0.06$\\
& & &1611.2005 &$ -2.8665$&ESI&$[ -80, 100]$&$<  42.6$&$< 15.32$& 
\enddata

\tablecomments{This table is available in its entirety in a machine-readable form in the online journal. A portion is shown here for guidance regarding its form and content.}
\tablenotetext{a}{Velocity interval over which the equivalent width and
column density are measured.}
\tablenotetext{b}{Rest equivalent width.}
\end{deluxetable*}

%% file: figset17.tex
\figsetstart
\figsetnum{17}
\figsettitle{Velocity profiles of DLAs.}

\begin{figure*}
\figurenum{17}
\epsscale{0.7}
\plotone{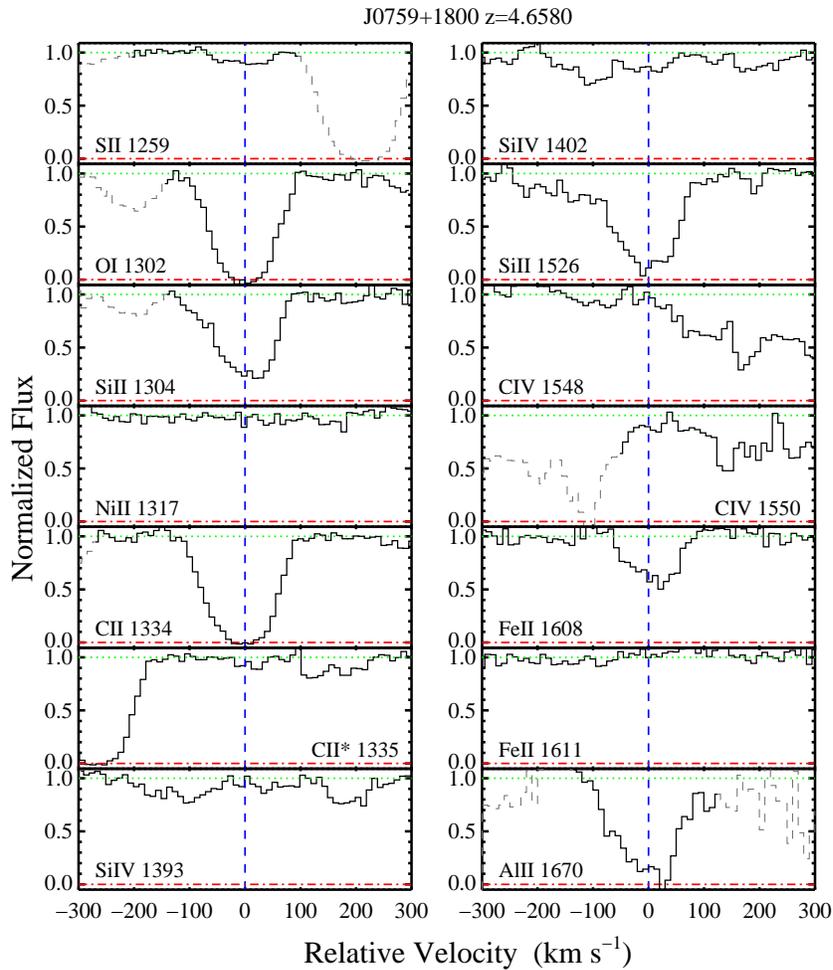}
\caption{Velocity profiles of metal transitions for each DLA in Table \ref{metal:tab1}. These figures are available in the online Figure set. }
\end{figure*}